\documentclass[acmsmall,screen]{acmart}

\usepackage{booktabs} 
\usepackage{color}
\usepackage{listings}
\usepackage{enumitem}
\usepackage{threeparttable}
\usepackage{multirow}
\usepackage{hhline}
\usepackage{subfigure}
\usepackage{array}
\usepackage{pifont}
\usepackage{booktabs}
\usepackage{xcolor}
\usepackage{xspace}
\usepackage[vlined, ruled, linesnumbered]{algorithm2e}
\usepackage{graphics}
\usepackage{epsfig}
\usepackage{mathrsfs} 

\usepackage{tcolorbox} 
\usepackage{tablefootnote}

\usepackage{cleveref} 

\usepackage{hyperref}
\usepackage{amsmath,amsfonts}
\usepackage{bbding}

\usepackage{wrapfig}

\usepackage{colortbl}
\usepackage[normalem]{ulem}
\useunder{\uline}{\ul}{}

\usepackage{graphicx}

\usepackage[normalem]{ulem}

\definecolor{codegreen}{rgb}{0,0.6,0}
\definecolor{codegray}{rgb}{0.5,0.5,0.5}
\definecolor{codepurple}{rgb}{0.58,0,0.82}
\definecolor{backcolour}{rgb}{0.95,0.95,0.92}

\lstdefinelanguage{DL}{
  classoffset=1,
  morekeywords={Flatten,Split},
  keywordstyle=\color{blue},
  classoffset=2,
  morekeywords={InputTensor},
  keywordstyle=\color{purple},
  classoffset=3,
  morekeywords={shape, axis, sections, split},
  keywordstyle=\bfseries,
}

\newcommand{\pre}[1]{}

\newcommand{\rev}[1]{#1}

\newcommand{\CodeIn}[1]{{\small\texttt{#1}}}

\newcommand{\toolname}{GReduce\xspace}

\newcommand{\ti}{Validity-Preserving Delta Debugging via Generator \rev{Trace Reduction}\xspace}

\newcommand{\tz}{\emph{Trace Instrumentation\xspace}}
\newcommand{\tone}{\emph{Trace Reduction\xspace}}
\newcommand{\ttwo}{\emph{Trace-Aligned Re-Execution\xspace}}
\newcommand{\rawtz}{Trace Instrumentation\xspace}
\newcommand{\rawtone}{Trace Reduction\xspace}
\newcommand{\rawttwo}{Trace-Aligned Re-Execution\xspace}

\newcommand{\rtt}{reduced trace\xspace}
\newcommand{\rtts}{reduced traces\xspace}
\newcommand{\str}{sequence-based trace reduction\xspace}
\newcommand{\ttr}{tree-based trace reduction\xspace}
\newcommand{\inft}{\emph{infeasible trace alignment\xspace}}
\newcommand{\iinft}{\emph{Infeasible trace alignment\xspace}}
\newcommand{\redq}{$Size \slash Size_{o}$}

\newcommand{\Luyao}[1]{{\color{blue} \bf \{Luyao: {#1}\}}}

\newcommand{\gJS}{JavaScript Program Generator}

\newcommand{\smalltitle}[1]{\noindent\textbf{#1}}
\DeclareRobustCommand\longtwoheadrightarrow {\relbar\joinrel\twoheadrightarrow}

\newcommand{\szero}{\emph{halt}}
\newcommand{\sone}{\emph{bypass}}
\newcommand{\stwo}{\emph{re-align}}

\newcommand{\ssa}{h}
\newcommand{\ssb}{b}
\newcommand{\ssc}{r}
\newcommand{\da}{S}
\newcommand{\db}{T}

\newcommand{\genin}{\CodeIn{"\color{red}a\color{orange}b\color{green}c\color{blue}\textbackslash n\color{red}a\color{orange}b\color{green}c\color{blue}\textbackslash n\color{black}"}}
\newcommand{\lf}{{\color{blue}{\textbackslash n}}}
\newcommand{\ca}{\color{red}a\color{black}}
\newcommand{\cb}{\color{orange}b\color{black}}
\newcommand{\cc}{\color{green}c\color{black}}
\newcommand{\cn}{\color{blue}\textbackslash n\color{black}}

\definecolor{codegreen}{rgb}{0,0.6,0}
\definecolor{codegray}{rgb}{0.5,0.5,0.5}
\definecolor{codepurple}{rgb}{0.58,0,0.82}
\definecolor{backcolour}{rgb}{1,1,1}
 
\lstdefinestyle{mystyle}{
    backgroundcolor=\color{backcolour},   
    commentstyle=\color{codegreen},
    keywordstyle=\color{magenta},
    numberstyle=\tiny\color{codegray},
    stringstyle=\color{codepurple},
    basicstyle=\ttfamily\footnotesize,
    breakatwhitespace=false,         
    breaklines=true,                 
    captionpos=b,                    
    keepspaces=true,                 
    numbers=left,                    
    numbersep=5pt,                  
    showspaces=false,                
    showstringspaces=false,
    showtabs=false,                  
    tabsize=2
}
 
\AtBeginDocument{%
  \providecommand\BibTeX{{%
    \normalfont B\kern-0.5em{\scshape i\kern-0.25em b}\kern-0.8em\TeX}}}

\copyrightyear{2024}
\acmYear{2024}
\setcopyright{rightsretained}

\begin{document}

\title{\ti}

\author{Luyao Ren}
\affiliation{%
    \institution{Key Laboratory of High Confidence Software Technologies (Peking University), Ministry of Education; School of Computer Science, Peking University, Beijing}
  \country{China}
}
\author{Xing Zhang}
\affiliation{%
    \institution{Key Laboratory of High Confidence Software Technologies (Peking University), Ministry of Education; School of Computer Science, Peking University, Beijing}
  \country{China}
}
\author{Ziyue Hua}
\affiliation{%
    \institution{Key Laboratory of High Confidence Software Technologies (Peking University), Ministry of Education; School of Computer Science, Peking University, Beijing}
  \country{China}
}
\author{Yanyan Jiang}
\affiliation{%
  \institution{State Key Laboratory for Novel Software Technology and Department of Computer
Science and Technology, Nanjing University}
  \country{China}
}
\author{Xiao He}
\affiliation{%
  \institution{School of Computer and Communication Engineering, University of Science and Technology Beijing}
  \country{China}
}
\author{Yingfei Xiong}
\affiliation{%
  \institution{Key Laboratory of High Confidence Software Technologies (Peking University), Ministry of Education; School of Computer Science, Peking University, Beijing}
  \country{China}
}
\author{Tao Xie}
\affiliation{%
  \institution{Key Laboratory of High Confidence Software Technologies (Peking University), Ministry of Education; School of Computer Science, Peking University, Beijing}
  \country{China}
}
\authornote{Corresponding author}



\renewcommand{\shortauthors}{}

\begin{abstract}
Reducing test inputs that trigger bugs is crucial for efficient debugging. Delta debugging is the most popular approach for this purpose. When test inputs need to conform to certain specifications, existing delta debugging practice encounters a validity problem: it blindly applies reduction rules, producing a large number of invalid test inputs that do not satisfy the required specifications.
This overall diminishing effectiveness and efficiency becomes even more pronounced when the specifications extend beyond syntactical structures.
Our key insight is that we should leverage input generators, which are aware of these specifications, to generate valid reduced inputs, rather than straightforwardly performing reduction on test inputs.
In this paper, we propose a generator-based delta debugging method, namely \toolname{}, which derives validity-preserving reducers.
Specifically, given a generator and its execution, demonstrating how the bug-inducing test input is generated,
\toolname{} searches for other executions on the generator that yield reduced, valid test inputs.
\rev{
The evaluation results on five benchmarks (i.e., graphs, DL models, JavaScript programs, SymPy, and algebraic data types) show that \toolname{} substantially outperforms state-of-the-art syntax-based reducers including Perses and T-PDD, and also outperforms QuickCheck, SmartCheck, as well as the state-of-the-art choice-sequence-based reducer Hypothesis, demonstrating the effectiveness, efficiency, and
versatility of \toolname{}.
}

\end{abstract}



\maketitle


\section{Introduction}\label{sec:s1}

In software testing, test inputs are often ~\emph{lengthy} when they are randomly generated. For example, in compiler testing, test inputs, generated by random testing tools or fuzzers, such as
Zest~\cite{padhye2019semantic} and Csmith~\cite{DBLP:conf/pldi/YangCER11}, tend to be large, complex, and messy. This complexity hampers developers' ability to reason and debug effectively, and the GCC bug reporting instructions~\cite{GCCDoc} explicitly asks for ``the preprocessed version of the file that triggers the bug.''
The developers of Csmith also acknowledge that bug-finding was most effective when the average size of random programs was 81 KB, which is largely out of necessity~\cite{DBLP:conf/pldi/YangCER11, CReduce}. 

Test input reduction is essential for software debugging to assist developers in identifying the root causes of bugs, especially when test inputs are lengthy. Test input reduction aims at simplifying a given test input, usually by reducing its size.
Delta debugging~\cite{ODD, DD} is currently the most widely used technique for reducing the sizes of test inputs in practice.
Given a test input and a specific property it exhibits (such as triggering an internal crash in the software under testing), delta debugging greedily finds a minimal version of the input that still satisfies the property.
DD algorithm~\cite{DD}, also named ~\emph{ddmin}, is the first algorithm that conducts an iterative search of manipulations on the given test input.
This algorithm inspired many follow-up approaches, such as HDD~\cite{HDD} that manipulates the tree structures of the test inputs parsed from a predefined grammar.


Unfortunately, delta-debugging algorithms are \emph{speculative} in nature and may generate a large number of \emph{invalid} test inputs during the reduction procedure.
The \textit{ddmin} family of algorithms waste time in examining test inputs that cannot manifest the bug (i.e., ineffectiveness) or enumerating invalid test inputs (i.e., inefficiency).
Some existing delta debugging practice, such as Perses~\cite{Perses}, utilizes predefined grammar of test inputs to conduct effective reductions, ensuring the syntactic correctness of the generated inputs.
However, the reduction based on the syntax rules has limitations when the specifications of test inputs become complex exceeding mere syntactical structures, such as
use-after-definition\footnote{Programming entities, such as variables and functions, must be correctly defined or declared before they are used (referenced or called) within the program.} and checksum\footnote{A checksum is a value calculated from a data set for the purpose of verifying the integrity and authenticity of that data. In many data formats, data should conform to a checksum specification; otherwise, internal exceptions may arise.}.
Alternatively, one could manually implement domain-specific reduction rules based on the software specifications, such as C-Reduce~\cite{CReduce} for the C programs, JSDelta~\cite{JSDelta} for JavaScript programs, ddSMT~\cite{niemetz2013ddsmt} for SMT formulas. However, it is evidenced that this manner requires considerable manual efforts.

This paper observes that \emph{lengthy} test inputs are often crafted by domain-specific test input generators.
Specifically, in addition to unit testing, software systems are usually stress-tested under a large number of randomly generated tests, and this practice is known as the generator-based testing ~\cite{DBLP:conf/pldi/YangCER11, 10.1145/3428264, padhye2019semantic, DBLP:journals/corr/abs-2302-00842, DBLP:conf/sp/WangCWL17, DBLP:conf/issta/HuaLRLZJ023}.
An interesting question naturally raises: \emph{can generators help reducing test inputs}?

This paper presents the first positive response to this question by proposing \toolname, a new approach to conducting optimistically validity-preserving reduction on test inputs.
When given the process of how a test input is crafted by a generator (as we called an \emph{execution} of the generator),
we try to reduce the potentially reducible subparts in the execution, e.g., generated random numbers and loop iterators, rather than blindly applying reduction rules to the original input. 
To achieve this goal, our approach removes subparts of the given execution while making best efforts to preserve the semantics of remaining part, via the following three steps:

\begin{enumerate}
    \item Disassemble the execution into a well-structured representation, referred to as a trace;
    \item Attempt to remove subparts of the trace to obtain a reduced trace;
    \item Re-execute the generator by aligning the execution with the reduced trace.
\end{enumerate}
As a result, our approach enables to obtain reduced test inputs by ``similar'' but smaller generator's execution, to finally obtain the reduction result effectively and efficiently.


To demonstrate \toolname{} as a practical approach for reducing the bug-inducing test inputs,
we employ \toolname{} on existing generators and empirically evaluate its effectiveness and efficiency in test input reduction \pre{in three distinct domains: graphs, deep learning (DL) models, and JavaScript programs}\rev{on five benchmarks: graphs in Zest~\cite{padhye2019semantic}, deep learning (DL) models in Isra~\cite{DBLP:journals/corr/abs-2302-00842}, JavaScript programs in Zest~\cite{padhye2019semantic}, SymPy benchmark~\cite{DBLP:conf/ecoop/MaciverD19}, and SmartCheck benchmark~\cite{DBLP:conf/ecoop/MaciverD19, pike2014smartcheck}}.
Our evaluation results show that \toolname{} outperforms state-of-the-art reducers.%
\pre{with (1) smaller results: \toolname{}'s reduction result is 28.5\%, 34.6\%, 75.6\% in size of those from Perses on graphs, DL models, and JavaScript programs; (2) shorter reduction time: \toolname{} takes 17.5\%, 0.6\%, 65.4\% time taken by Perses on graphs, DL models, and JavaScript programs respectively
}%
\rev{ For benchmarks of graphs, DL models, and JavaScript programs, \toolname{}'s reduction result is 28.5\%, 34.6\%, 75.6\% in size of those from Perses~\cite{Perses}, and \toolname{} takes 17.5\%, 0.6\%, 65.4\% time taken by Perses, respectively;
\toolname{}'s reduction result is 15.2\%, 8.8\%, 83.9\% in size of those from T-PDD~\cite{wang2023probabilistic}, and \toolname{} takes 13.9\%, 0.5\%, 22.0\% time taken by T-PDD, respectively.
For SymPy and SmartCheck benchmarks from Hypothesis~\cite{DBLP:conf/ecoop/MaciverD19}, GReduce takes much less reduction time while achieving better or comparable reduction results compared to Hypothesis~\cite{DBLP:conf/ecoop/MaciverD19}, QuickCheck~\cite{DBLP:conf/icfp/ClaessenH00} and SmartCheck~\cite{pike2014smartcheck}}, highlighting the effectiveness, efficiency and versatility of \toolname{}.

In summary, this paper makes the following contributions:
\begin{itemize}
    \item A new perspective: we introduce a fresh viewpoint to resolve the validity problem in delta debugging by deriving validity-preserving reducers from input generators. 
    \item An innovative approach: we propose a novel approach for conducting validity-preserving delta debugging via test input generators, named \toolname{}. Given a test input generator as well as an execution on the generator which yields a bug-inducing input,
    \toolname{} conducts reduction on the execution to yield reduced test inputs by removing subparts of the given execution while making best efforts to preserve the semantics of remaining part.
    \item Implementations and evaluations: we demonstrate the effectiveness, efficiency, and versatility of our approach through implementations and evaluations \pre{in three different domains}\rev{on five benchmarks across distinct domains}.

    
    
\end{itemize}




\section{Background and Motivation}\label{sec:s2}



In this section, we first present a motivating example to illustrate the background of the problem: under the setting of generator-based testing, how existing delta debugging practice \pre{in fall}\rev{falls} short on the motivating example. Subsequently, we present the main idea of our approach and illustrate how it addresses the problem, using the same motivating example for demonstration.


\subsection{Motivating Example: Password Validation}

Imagine a Web application that asks a user to enter the password twice in the sign-up form to avoid mistyping. The page will submit the form to the backend to create a user account only when the two passwords pass an input validation---they must be identical and consist solely of lowercase letters.

\begin{wrapfigure}{r}{0.27\textwidth}
  \centering
  \includegraphics[width=\linewidth]{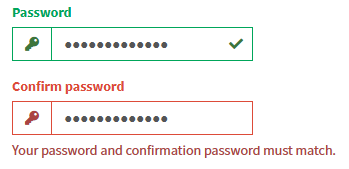}
    
\end{wrapfigure}

Suppose that the passwords are sent to the application line by line, with each password followed by a line break ``$\lf$''. The application only further processes the valid inputs that comply the following \emph{input specification}\footnote{$*$ is the Kleene star, $S^{*}$ means the set of all strings over symbols in $S$ (including the empty string).}:
$$\mathbb{L}=
\big\{
  ww |\, w \in \{a, ..., z\}^{*}\lf\color{black}
\big\}.$$%
Assume that the application has a bug that is triggered when the passwords pass the input validation and contain the character ``\CodeIn{\cc}''.
This example case resembles real-world bugs that require 
specific character combinations (e.g., an emoji) to trigger. 
Note that we employ this motivating example throughout our discussions in Section~\ref{sec:s2}.

\subsection{Bug Finding and Debugging}

\smalltitle{Finding bugs with a test input generator}.
In real-world software systems, input specifications can be considerably complex. Generating \emph{valid} inputs can be challenging.
For example, if we feed the password-validation application with randomly generated strings as test inputs, it is unlikely that these inputs will conform to the input specifications (e.g., $\mathbb{L}$). Consequently, these test inputs will be rejected at an early stage within the application logic.

To generate valid test inputs effectively, the technique of \emph{generator-based testing} has been proposed.
Generator-based testing utilizes the \emph{domain-specific generator} that algorithmically realizes input specifications  to construct ~\emph{valid} test inputs.
For example, Csmith~\cite{DBLP:conf/pldi/YangCER11} can generate C programs that meet the standard of C language and do not contain undefined behaviors.
Due to its high effectiveness in generating valid inputs, generator-based testing has shown great practice value, especially for testing software systems with complex input specifications~\cite{DBLP:conf/pldi/YangCER11, 10.1145/3428264, padhye2019semantic, DBLP:journals/corr/abs-2302-00842, DBLP:conf/sp/WangCWL17, DBLP:conf/issta/HuaLRLZJ023}.


In our motivating example, we can test the backend logic with valid test inputs generated by the following generator written in Python:


\begin{lstlisting}[xleftmargin=4em, language=Python, caption=A generator of generating random strings in language L, label={lst:g0}]
import random
from string import ascii_lowercase

def gen(seed=26524): # This default seed is for generating "abc\nabc\n".
    random.seed(seed)
    w = ""
    n = random.choice(range(20)) # Set the number of loop iterations.
    for i in range(n):
        w += random.choice(ascii_lowercase) # Append a letter into w.
    w += '\n'
    return w + w

print(gen())
\end{lstlisting}

This generator constructs test inputs as follows: it first concatenates multiple random characters to form a string (Lines 7--9), where the length of the string is bounded by a constant; afterwards, it appends a line break at the end of the string (Line 10); finally, it duplicates the string sequentially as the generated test input (Line 11).
With \texttt{seed=26524}, the generator returns a buggy input that contains character  ``$\CodeIn{\cc}$'':
$$I=\genin{}$$

\smalltitle{Delta Debugging}.
While input generators are very good at \emph{triggering} bugs,
the resulting buggy test inputs may be excessively long.
This lengthiness can make it extremely difficult in pinpointing the root causes.
To ease debugging, test input reduction is essential in practice.

\newcommand{\ddmin}{\emph{ddmin}\xspace}

\begin{figure}[htbp]
	\centering
	\begin{minipage}{0.55\linewidth}
		\centering
		\includegraphics[width=0.95\linewidth]{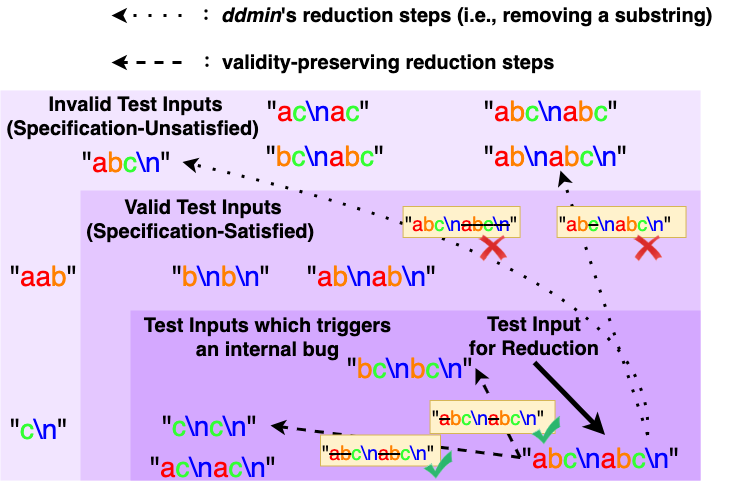}
		\caption{Reduction steps and derived test inputs.}
    \label{fig:space}
	\end{minipage}
        \begin{minipage}{0.44\linewidth}
\begin{algorithm}[H]
\small

\caption{The \emph{ddmin} algorithm}
\label{alg:dd}

\KwIn{$I$ is the test input for reduction; $P$ checks whether input $I$ is buggy.}
\KwOut{the test input after reduction.}

\textbf{function} $\ddmin(I, P)$ \\
\Begin{
    \For{substring $I[i:j] \subseteq I$}
    {
        $I' \leftarrow I \setminus I[i:j]$\; 
        \If{$P(I')$}{
            $I \leftarrow I'$\;
        }
    }
    \Return{I}\;
}

\end{algorithm}
	\end{minipage}
\end{figure}

Observing that \emph{not all parts in the test input contribute equally to bug manifestation}, the seminal delta debugging work \ddmin \cite{DD}, as shown in \Cref{alg:dd}, greedily attempts to remove consecutive substrings from the original test input to derive a smaller yet buggy input.
The progress of removing a substring is referred to as a \emph{reduction step} (Line 4 in \Cref{alg:dd}).
These reduction steps can be made more effective by taking the grammar of the original input into account \cite{HDD, Perses, 10.1145/3586049}. This approach allows  original input to be converted into a syntax tree, enabling  the precise removal of local constructs, such as matched parentheses.




Unfortunately, our bug-inducing test input $\genin{}$
is \emph{irreducible} for existing delta-debugging techniques:
any attempt to reduce $I$ to $I'$ by removing a substring does not enforce the validity of the reduced input, as shown in Figure~\ref{fig:space}.
Consequently, $I'$ becomes an invalid input (i.e., $I' \notin \mathbb{L}$) and is unable to trigger the bug.
To address this, developers may define some customized reduction rules 
that explicitly specify possible (and potentially profitable) validity-preserving reduction steps,
e.g., removing $I[i]$ and $I[i+(|I|\slash 2)]$ simultaneously.
It is worth noting that such a validity-preserving reducer can be automatically derived from an input generator.

\subsection{Test Input Reduction with Test Input Generator}

The common case in delta debugging is that we lack domain-specific reduction tools that preserve the validity, but we do have a generator that ensures the validity of generated test inputs. 
Thus, why not hack the generator to obtain a validity-preserving reducer?

Consider the trace of execution on the generator with \texttt{seed=26524}, which illustrates how the generator crafts the test input $\genin{}$. The trace is represented as a sequence of operations as follows (the formal definition of traces, executions, and operations are shown in Section~\ref{s3.1}):
\begin{equation*}\label{eq:trace1}
    T: \langle \CodeIn{w = ""}, ..., \CodeIn{w += "\ca"}, ..., \CodeIn{w += "\cb"}, ..., \CodeIn{w += "\cc"}, \CodeIn{w += "\cn"}, \CodeIn{w = w + w}, \CodeIn{return w} \rangle.
\end{equation*}
An immediate observation is that ``removing'' some loop iterations from the loop yields smaller test inputs.
This paper proposes to ``remove'' \emph{consecutive segments of the execution's trace} (instead of test input string) in a reduction step. In our example, we attempt to remove the first two loop iterations:
\begin{equation*}\label{eq:trace2} 
T^{\dagger}: \langle \CodeIn{w = ""}, ..., \mbox{\sout{\CodeIn{w += "\ca"}}}, ..., \mbox{\sout{\CodeIn{w += "\cb"}}}, ..., \CodeIn{w += "\cc"}, \CodeIn{w += "\cn"}, \CodeIn{w = w + w}, \CodeIn{return w} \rangle.
\end{equation*}
If we could re-execute the generator to align with above trace $T^{\dagger}$, we would obtain a smaller test input
$I^{\prime}=\CodeIn{"\cc\cn\cc\cn"}$.
We generalize this idea as our approach, named \toolname{}. Further details are presented in the rest of this paper.

\section{Approach}\label{sec:s3}


We first establish definitions of the terminologies and symbols introduced by our approach in Section~\ref{s3.1}, as an important preliminary to \pre{starting}\rev{start} the detailed approach.
We then demonstrate an overview of our approach in Section~\ref{s3.2}.
The remaining portions of this section delve into the details of three components in our approach: \tz{} (in Section~\ref{s3.3}), \tone{} (in Section~\ref{tone}) and \ttwo{} (in Section~\ref{ttwo}).

\subsection{Preliminary: Terminologies and Symbols}\label{s3.1}
In this section, we give definitions of terminologies and symbols used by our approach. 
We also give the definitions of delta debugging for demonstrating the difference between our approach and the established practice of delta debugging.






\begin{definition}[Generator]
A generator $G$ is a program that relies on a source of randomness, usually a pseudo-random generator that can be configured with a seed.
Given different seeds, $G$ generates various test inputs, following the principles of generator-based testing, as seen in tools such as QuickCheck~\cite{DBLP:conf/icfp/ClaessenH00} and Zest~\cite{padhye2019semantic}.
\end{definition}




\begin{definition}[State and Operation]
Any program, including the generators, is essentially a state transition system in which a state $S$ refers to the snapshot of all runtime information (variable values, program counter, and memory contents) at a program's execution point.
A program can perform an operation $O$ for state transition $S
\stackrel{O}\longrightarrow S^{\prime}$. 
This transition typically occurs by interpreting a program’s subpart denoted as $Prog(O)$ (e.g., a statement, an instrument, or a subroutine).
\end{definition}




\begin{definition}[Execution]
An execution $E$ on a program $G$ is modeled as a sequence of transitions defined over states:
$
E:
S_1
\stackrel{O_1}\longrightarrow S_2
\stackrel{O_2}\longrightarrow S_3
\stackrel{O_3}\longrightarrow S_4
\stackrel{O_4}\longrightarrow ...
\stackrel{O_n}\longrightarrow S_{n+1}.
$
\end{definition}


\begin{definition}[Trace]
One can instrument the program to obtain its execution trace $T$,
an operation sequence $\langle O_i \rangle$ of an execution $E$, i.e., $T: \langle O_1, O_2, O_3, ... , O_{n}\rangle.$
\end{definition}

\begin{definition}[Delta Debugging]
Given a test input $I$ and a property that $I$ exhibits, the goal of delta debugging is to explore reduction candidates $I'$, s.t., $I^{\prime} \preceq I$, and identify a minimal $I^{*}$ that still exhibits the property, as depicted in ~\Cref{eq1}.
The property is defined as a function $P$, where $P(I)=\mbox{True}$ if $I$ exhibits the property such as triggering a bug of interest.

\begin{equation}\label{eq1} 
I^{*} = \arg\min \limits_{\displaystyle I^{\prime} \preceq I}^{} |I'| \ s.t. \ P(I') = \mbox{True}.
\end{equation}
\end{definition}

Exploring the design space of the relation $\preceq$ yields a spectrum of delta debugging techniques.
The earliest delta debugging algorithm, \emph{ddmin}~\cite{ODD, DD}, regards a test input $I$ as a linear sequence and searches subsequences of $I$ with a greedy algorithm~\footnote{
Finding the globally minimal one is usually infeasible since hitting set problem is known to be NP-complete~\cite{DBLP:conf/coco/Karp72}.
Instead, ~\emph{ddmin} algorithm ensures the result is 1-minimal: if deleting any single element makes the result lose the property.}:
$I^{\prime} \preceq I \iff I^{\prime} = \mbox{a subsequence of } I.$
Based on \emph{ddmin}, Hierarchical Delta Debugging (HDD) ~\cite{HDD} parses an test input as a parse tree $T$ and applies ~\emph{ddmin} on every level of $T$: 
$I^{\prime} \preceq I \iff 
T^{\prime}$ = a subtree\footnote{Here, a subtree refers to a subgraph of the given tree with keeping the root.} of $T$, where $T$ is a parse tree of $I$, $I^{\prime}$ is an unparsing ~\footnote{Unparsing, also named pretty printing~\cite{rendel2010invertible}, refers to constructing the result from a parse tree, i.e., performing the reverse operation of parsing, e.g., traversing the parse tree of a program to print the code of the program.} result of $T^{\prime}$.

\subsection{Approach Overview}\label{s3.2}

With a generator, any generated test input $I$ is associated with its execution $E$.
In other words, the execution $E$ yields the test input $I$, written as $E \rightsquigarrow I$.
Instead of conducting reduction directly on $I$,
\toolname reduces $E$ to obtain other executions that yield reduced test inputs such as $E^{\prime} \rightsquigarrow I^{\prime}$:
$$
I^{*} = GReduce(G, E, P)
    = \arg\min \limits_{\displaystyle E^{\prime} \preceq E, E^{\prime}
    \rightsquigarrow I^{\prime}}^{} |I'| \ s.t. \ P(I') = \mbox{True}.
$$

\begin{algorithm}
\caption{The framework of GReduce}
\label{alg:GReduce}

\KwIn{$G$: a generator; $E$: an execution of the generator which yield the test input for reduction; $P$: a property test function.}
\KwOut{the test input after reduction.}

\textbf{function} $\toolname{}(G, E, P)$ \\
    \Begin{
    

    $T \leftarrow \mbox{\rawtz{}}(E)$\;
    \For{$T^{\dagger} = \mbox{\tone{}}(T)$} 
    {
        $E^{\prime} \leftarrow \mbox{\rawttwo{}}(G, T^{\dagger})$\;
        $I^{\prime} \leftarrow Output(E^{\prime})$\;
        \If{$P(I^{\prime})$} 
        {
            $T \leftarrow T^{\dagger}$\;
        }
    }
    $I^{*} \leftarrow Output(\mbox{\rawttwo{}}(G, T))$\;
    \Return{$I^{*}$}\;
}
\end{algorithm}

Specifically, \toolname{} tries to \emph{remove subparts of an execution while making best efforts to preserve the semantics of remaining parts}. To achieve it, we design three components in \toolname{}:
\begin{enumerate}
    \item \tz{} disassembles a given execution $E$ on the generator into a trace $T$;
    \item \tone{} takes a trace $T$ as input and outputs \rtts{} such as $T^{\dagger}$;
    \item \ttwo{} takes a generator $G$ as well as a \rtt{} $T^{\dagger}$ as inputs, and re-executes the generator $G$ with the goal of aligning the trace of re-execution, $T^{\prime}$, with the given trace $T^{\dagger}$,
finally yielding a generated test input $I^{\prime}$.

\end{enumerate}

\pre{
With these three components, \toolname{} is able to conduct reduction on the executions:
$$
    E^{\prime} \preceq E
    \Longleftrightarrow
    E^{\prime} =
    \mbox{\ttwo{}}(G, \mbox{\tone{}}(\mbox{\tz{}}(E))).
$$}
\rev{}
The \rev{overall} framework of \toolname{} is shown in Algorithm~\ref{alg:GReduce}. Given a generator $G$ as well as an execution $E$ on the generator (which yields a test input $I$), \toolname{} first instruments $E$ to obtain a trace $T$ by ~\tz{} (Line 3), then, \toolname{} iteratively derives \rtts{} from the trace $T$ by \tone{} (Line 4) until finding out an optimal trace.
In each iteration, \toolname{} re-executes the generator to obtain executions that yield reduced test inputs by \ttwo{} (Lines 5-6), until the optimal trace is found (Line 9). 
Note, we denote $Output(E)$ as the test input generated by execution $E$ in Algorithm~\ref{alg:GReduce}, i.e., $E \rightsquigarrow Output(E)$.

An overview of ~\toolname{} is shown in ~\Cref{fig:workflow}. The details of three components in \toolname{} are described in the following sections.




\subsection{\rawtz{}: Disassembling the Execution into a Trace}~\label{s3.3}

Executing a generator $G$ with a specific source of randomness (e.g., a pseudo-random generator with a fixed seed) makes it deterministic, yielding an execution 
$$
E: S_1
\stackrel{O_1}\longrightarrow S_2
\stackrel{O_2}\longrightarrow S_3
\stackrel{O_3}\longrightarrow S_4
\stackrel{O_4}\longrightarrow ...
\stackrel{O_n}\longrightarrow S_{n+1}
$$
where we can collect trace
$$T: \langle O_1, O_2, O_3, ... , O_{n}\rangle$$
by instrumenting~\cite{ins} the generator.
Each operation $O_i$ in the trace corresponds to a subpart of execution that transforms the program state through a generator's subpart $Prog(O_i)$.
The nature of $Prog(O_i)$ depends on the granularity of instrumentation, which could be a statement, an instrument, or a subroutine.

\begin{figure}[]
    \includegraphics[width=0.92\linewidth]{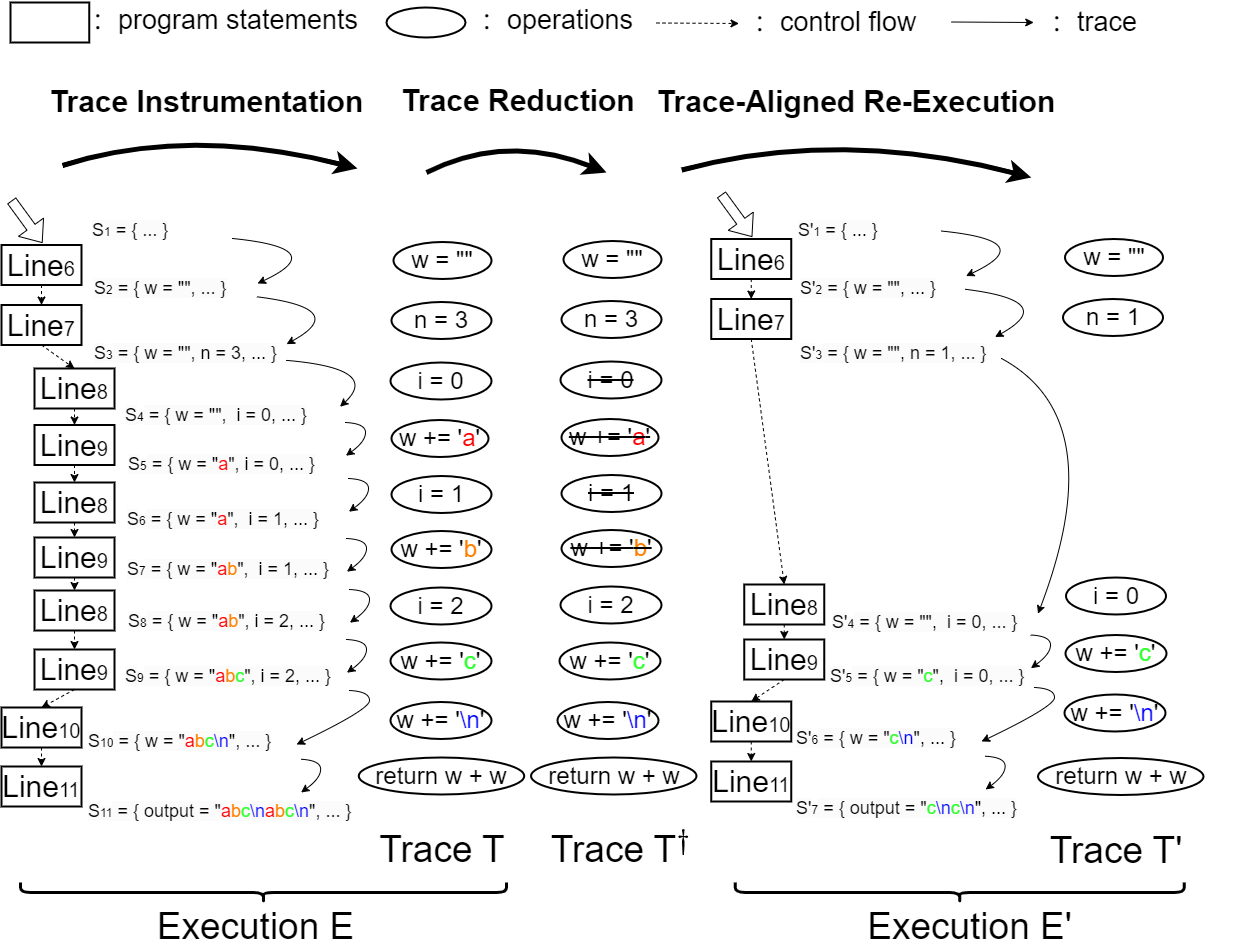}
    \caption{The overall workflow demonstrating how ~\toolname{} finds a reduced generated inputs by aligning the trace of re-execution \( T^{\prime} \) with the original given trace \( T \).}
    \label{fig:workflow}
\end{figure}

For example, statement-level instrumentation of the program in Listing~\ref{lst:g0} results in the trace (also shown in Figure~\ref{fig:workflow}):
\begin{equation}~\label{eq:trace}
    \begin{aligned}
        T: \langle
        &
        O_1: \CodeIn{w = ""} ,
        O_2: \CodeIn{n = 3} ,
        O_3: \CodeIn{i = 0} ,
        O_4: \CodeIn{w += "\color{red}a\color{black}"} ,
        O_5: \CodeIn{i = 1} ,
        O_6: \CodeIn{w += "\color{orange}b\color{black}"} , \\&
        O_7: \CodeIn{i = 2} ,
        O_8: \CodeIn{w += "\color{green}c\color{black}"} ,
        O_9: \CodeIn{w += "\color{blue}\textbackslash n\color{black}"} ,
        O_{10}: \CodeIn{return w + w}
        \rangle.
    \end{aligned}
\end{equation}








\subsection{\rawtone{}: Labeling Subparts of the Given Trace for Removal}\label{tone}
Given a trace $T$ (as the result of \tz{}), ~\tone{} labels ``removals'' on the operations of trace $T$ to derive ~\emph{\rtts}.
It primarily comprises two steps: initially recognizing subparts within the trace based on two patterns, which we refer to as ~\emph{reducible parts} of the trace; and subsequently, labeling ``removals'' on operations within these ~\emph{reducible parts}. If $S$ represents a subpart of the trace, $\mbox{\sout{$S$}}$ is denoted as labeling ``removals'' for all of operations within $S$. 

As an example of ~\emph{reducible parts}, for the generator in Listing~\ref{lst:g0}, 
in Line 7-9, the loop in the generator may create abundant number of elements in the generated test inputs.
    Not all elements contributed to triggering the bug.
    Because the number of loop iterations is determined by a random choice (i.e., \CodeIn{n = random.choice(range(20))} in Line 7), making this choice smaller as well as ``removing'' some of loop iterations (as if we manipulate randomness) results in producing reduced test inputs.

Based on our preliminary study of generators collected in Zest~\cite{padhye2019semantic} \rev{as well as the inspirations of several previous research work~\cite{Perses, vikram2021growing, DBLP:conf/ecoop/MaciverD19}}, we designed two heuristic patterns to recognize reducible parts.
We \rev{did a study on 15 generators in Zest~\cite{padhye2019semantic} and} found that
\pre{these two patterns are common in generators (appears in 13/15), and they cover the majority of random choices that affect the size of generated test inputs in generators.}
\rev{a significant 94.3\% (50 out of 53) of the random choices that alter the size of the execution trace occur within the predicates of loops and selections in the generator (the remaining 3 out of 53 are in 2 of the 15 studied generators).
Thus, we designed two heuristic patterns to identify the vast majority of cases where reduced traces could be derived.
Our designed patterns also reference the grammar normal forms in Perses~\cite{Perses}, the properties of generators studied in Bonsai Fuzzing~\cite{vikram2021growing}, and reduction strategies for choice sequences designed in Hypothesis~\cite{DBLP:conf/ecoop/MaciverD19}, as demonstrated in Section~\ref{sec:relatedwork} in detail.}
The evaluation in Section~\ref{sec:s5} also shows the effectiveness of these two heuristic patterns for reduction.



\pre{The following two patterns are employed in \tone:} \rev{The details of these two patterns are as follows:}

\textbf{Reducible Loop Pattern} refers to a loop whose iteration count is determined by a randomly chosen $n$ in a range $[0, N)$, where $N$ is a constant.
We denote such a pattern as $L\{n\}$. For example, consider the following Python code (assuming the loop iteratives exactly $n$ times without any premature termination, such as \CodeIn{break}/\CodeIn{return} inside the loop):
\begin{lstlisting}[language=Python, label={lst:loop1}]
    n = random.choice(range(N))
    for _ in range(n):
        {block}
    ...
\end{lstlisting}

At the initiation of $L\{n\}$, we denote $L_{init}$ as a special subpart of the trace, which represents a single operation of setting the number of loop iterations by a decision on a random choice. Next, to distinguish each iteration of the loop in $L\{n\}$, we denote the subpart of the trace for the i-th loop iteration as $L_{i}$. Thus, $L\{n\}$ is the concatenation~\footnote{If $X$ and $Y$ are subparts of the trace, i.e., sequences of operations, then $ \langle X,Y \rangle  $ denote a sequence of the concatenation of $X$ and $Y$. E.g., $X: \langle O_1 \rangle, Y: \langle O_2, O_3 \rangle$, $ \langle X,Y \rangle: \langle O_1, O_2, O_3 \rangle.$} of $L_{init}$ and all of $L_{i}$, represented as follows:
$$
T: \langle ..., L\{n\}, ... \rangle, L\{n\}: \langle L_{init}, L_{1}, L_{2}, L_{3}, ..., L_{n-1}, L_{n} \rangle.
$$

As an example, a reducible loop pattern in the execution's trace of the motivating example (i.e, Listing~\ref{lst:g0} in Section~\ref{sec:s2}) is as follows:
$$ L\{3\}:
\langle 
L_{init}: O_{2} (\CodeIn{n = 3}),
L_{1}: O_{3} O_{4} (..., \CodeIn{w += "\ca{}"}),
L_{2}: O_{5} O_{6} (..., \CodeIn{w += "\cb{}"}),
L_{3}: O_{7} O_{8} (..., \CodeIn{w += "\cc{}"})
\rangle.
$$

For a reducible loop pattern, e.g. a subpart $L\{n\}$ in the trace $T$, \tone{} labels on subsequences of elements in $\langle L_1, ..., L_{n} \rangle$, to derive \rtts such as $T^{\dagger}$ as follows:
\begin{equation*}
    \begin{aligned}
T^{\dagger}: \langle ..., 
L_{init},
..., \mbox{\sout{$L_{x_1 - 1}$}}, L_{x_1},  
..., \mbox{\sout{$L_{x_2 - 1}$}}, L_{x_{2}},  
..., \mbox{\sout{$L_{x_k - 1}$}},
L_{x_{k}},
\mbox{\sout{$L_{x_{k} + 1}$}}, 
...
\rangle
,
where\ 1 \leq x_1 < x_2 < ... < x_{k} \leq n.
    \end{aligned}
\end{equation*}

\textbf{Reducible Selection Pattern} refers a program's subpart that is conditionally executed on a random choice like:
\begin{lstlisting}[language=Python, label={lst:choice}]
    if (random.choice([False, True])):
        {block}
    ...
\end{lstlisting}


We denote \pre{$S_{init}$}\rev{$B_{init}$} as a special subpart of the trace, which is the predicate of the execution on a program’s subpart (e.g. \CodeIn{\{block\}} in above example). \pre{$S_{init}$}\rev{$B_{init}$} is a single operation of a decision on a random choice which randomly returns \CodeIn{True} or \CodeIn{False}. If the decision is \CodeIn{True}, the program’s subpart will be executed, and the corresponding subpart of the trace is denoted as \pre{$S_{block}$}\rev{$B_{block}$}.

We denote such pattern as \pre{$S?$}\rev{$B?$} when the decision in \pre{$S_{init}$}\rev{$B_{init}$} is \CodeIn{True} (otherwise, when the decision is \CodeIn{False}, we do not regard it as a reducible part in the trace),
and represent such pattern in the trace as follows:
\pre{$$T: \langle ..., S?, ... \rangle, S?: \langle S_{init}, S_{block} \rangle .$$}
\rev{$$T: \langle ..., B?, ... \rangle, B?: \langle B_{init}, B_{block} \rangle .$$}

For a reducible selection pattern, e.g., a subpart \pre{$S?$}\rev{$B?$} in the trace $T$, \tone{} could choose to label (or not label) on \pre{$S_{block}$}\rev{$B_{block}$} to derive one of following \rtts:
\pre{
$$T^{\dagger}:\langle ..., S_{init}, \mbox{\sout{$S_{block}$}}, ... \rangle
, or\ 
\langle ..., S_{init}, S_{block}, ... \rangle
. $$
}
\rev{
$$T^{\dagger}:\langle ..., B_{init}, \mbox{\sout{$B_{block}$}}, ... \rangle
, or\ 
\langle ..., B_{init}, B_{block}, ... \rangle
. $$
}

\rev{Back to the overall workflow of \tone{}, our approach will first identify reducible parts in the execution trace of the generator based on the above reducible patterns, then derive reduced traces for all of the reducible parts.

For the first step, our approach will conduct a pattern matching process through program analysis to identify those reducible parts. Explicitly, our approach first traverses the abstract syntax tree (AST) of generators by mature program analysis tools such as Python AST library~\cite{PyAST}. During the traversal, our approach identifies reducible patterns by comparing both syntax structures and contents of APIs, which produce random choices used within generators.
It is noteworthy that we did not support all kinds of generators because patterns vary across different implementations. For instance, by using different APIs that produce random choices in Python, \CodeIn{random.choice([0, 1])} could also be rewritten as \CodeIn{random.randint(0, 1)}. Currently, our implementation has supported common APIs in Java and Python, as a prototype of our approach for evaluation. As a supplement, if the APIs are not supported, we allow users to manually denote the random choices and reducible patterns by adding annotations in the generator.
}

\pre{To}\rev{For the second step, to} derive reduced traces, for all of subparts in the reducible parts (i.e., $L_{i}$ and \pre{$S_{block}$}\rev{$B_{block}$}), we need to decide whether to label ``removals'' or not. Each labeling scheme on the trace corresponds to a reduced trace. Enumerating all of reduced traces is a straight-forward method as shown in the naive implementation in Section~\ref{sec:s4.1}. However, it is unacceptable in practice due to its exponential complexity. To optimize it, rather than employing a brute-force enumeration, we utilize practical methods for deriving reduced traces. Our approach involves leveraging greedy search strategies from existing delta debugging algorithms such as ~\emph{ddmin}~\cite{DD} and HDD~\cite{HDD}, in our actual implementation. We left the details of the implementation of optimizations in Section~\ref{s4.2}.

\subsection{\rawttwo{}: Aligning the Trace of Execution with a Reduced Trace}~\label{ttwo}

Note that the reduced trace, $T^\dagger$, is entirely based on our speculation on trace reduction patterns.
We would like to find a new execution $E^{\prime}$ with trace $T^{\prime}$ which can be aligned with $T^{\dagger}$, refered to as ~\emph{trace alignment}.
Formally, 
given a generator $G$ as well as a \rtt $T^{\dagger}$, derived from $T: \langle O_1, O_2, ..., O_n \rangle$, i.e., $ T^{\dagger} = \mbox{\tone{}}(T)$, 
~\ttwo{} aims to find another execution $E^{\prime}$ with its trace $T^{\prime}$ to align with $T^{\dagger}$, as we called \emph{trace alignment}, denoted as $T^{\prime} \approx T^{\dagger}$ :
\begin{equation*}
    \begin{aligned}
T^{\prime} : &\langle \mathit{O}^{\prime}_{1}, \mathit{O}^{\prime}_{2}, ..., \mathit{O}_{m-1}^{\prime}, \mathit{O}_{m}^{\prime}
\rangle
\\\approx T^{\dagger}: 
&\langle
..., \mbox{\sout{$O_{x_1 - 1}$}}, O_{x_1},
..., \mbox{\sout{$O_{x_2 - 1}$}}, O_{x_{2}},
..., \mbox{\sout{$O_{x_k - 1}$}},
O_{x_{k}}, \mbox{\sout{$O_{x_{k} + 1}$}}, ...
\rangle
, where\ 1 \leq x_1 < x_2 < ... < x_{k} \leq n\\
    \end{aligned}
\end{equation*}



To achieve this, ~\ttwo{} attempts to align each operation in trace $T^{\prime}$ with an unlabeled operation in \rtt $T^{\dagger}$ in order. In other words, it attempts to build a one-to-one (bijective) mapping between operations in the trace $T^{\prime}$ and unlabeled operations in the \rtt $T^{\dagger}$ as follows \rev{($O_{i}^{\prime}$ denotes an operation in $T^{\prime}$ and $O_{x_i}$ denotes the mapping operation in $T^{\dagger}$)}: 
\begin{equation}~\label{eq:opalign}
    \begin{aligned}
    T^{\prime} \approx T^{\dagger} \Longleftrightarrow
        m = k, 
        \forall 1 \leq i \leq k, 
        O_{i}^{\prime} \approx O_{x_i}
    \end{aligned}
\end{equation}

As an example, given a \rtt{} from the motivating example (i.e., Listing~\ref{lst:g0}) as follows:
\begin{equation*}
    \begin{aligned}
T^{\dagger}:
&
\langle O_1,
L_{init}: O_2,
\mbox{\sout{$L_{1}: O_3 O_4$}},
\mbox{\sout{$L_{2}: O_5 O_6$}},
L_{3}: O_7 O_8
, ...
\rangle \\
= 
&
\langle
        O_1: \CodeIn{w = ""}, 
        O_2: \CodeIn{n = 3} , 
        \mbox{\sout{$O_3$}} ,
        \mbox{\sout{$O_4$}} ,
        \mbox{\sout{$O_5$}} ,
        \mbox{\sout{$O_6$}} ,
        O_7: \CodeIn{i = 2} ,
        O_8: \CodeIn{w += "\cc"},
        ...,
\rangle
.
    \end{aligned}
\end{equation*}

\ttwo{} aligns the re-execution's trace $T^{\prime}: \langle O_1^{\prime}, O_2^{\prime}, O_3^{\prime}, O_4^{\prime}, ...\rangle$ with $T^{\dagger}$ through achieving $O_1^{\prime} \approx O_1, O_2^{\prime} \approx O_2, O_3^{\prime} \approx O_7, O_4^{\prime} \approx O_8$ and so on:
\begin{equation}~\label{tp}
    \begin{aligned}
T^{\prime}:
\langle
        O_1^{\prime}: \CodeIn{w = ""}, 
        O_2^{\prime}: \CodeIn{n = 1} , 
        O_3^{\prime}: \CodeIn{i = 0} , 
        O_4^{\prime}: \CodeIn{w += "\cc"} ,
        ...,
\rangle.
    \end{aligned}
\end{equation}

For $O_2^{\prime}$, the initiation of a reducible loop pattern, \ttwo{} manipulates the randomness in $O_2^{\prime}$ to modify the return value of $\CodeIn{random.choice(range(20))}$ (Line 7) to be $1$, resulting in $O_2^{\prime}$ being $ \CodeIn{n = 1} $.
Furthermore, to achieve $O_4^{\prime} \approx O_8$, \ttwo{} manipulates randomness in $O_4^{\prime}$ as the same with $O_8$, denoted as $Dec(O_4^{\prime})=Dec(O_8)$ (the detailed explanation of $Dec$ is in the next section), as follows:
\begin{equation*}
    O_4^{\prime} \approx O_8
    \Longrightarrow 
    \begin{cases}
        Prog(O_4^{\prime}) = Prog(O_8): \CodeIn{w += random.choice(ascii\_lowercase}),\\
        Dec(O_4^{\prime}) = Dec(O_8): \CodeIn{random.choice(ascii\_lowercase)} \rightarrow \CodeIn{"\cc{}"} 
    \end{cases}
    \Longrightarrow 
    O_4^{\prime}: \CodeIn{w += "\cc{}"}
    .
\end{equation*}

Consequently, \ttwo{} finds an execution $E^{\prime}$ with its corresponding trace $T^{\prime}$ as shown in ~\Cref{tp}. Next, we elaborate on how \ttwo{} attempts to achieve trace alignment in detail.

Trace alignment is based on the observation that the non-deterministic behavior in the generator is fully determined by the randomness (e.g., \CodeIn{random} in our example).
Thus, there is a chance of creating aligned trace by manipulating the randomness for non-deterministic behavior.

The randomness determines all of random choices made during the execution of the generator (e.g. the return values of function calls of \CodeIn{random.choice(...)} in Listing~\ref{lst:g0}).
To simplify (without losing generality), we denote a random choice as a non-deterministic function that takes a set as input and randomly picks one of elements in the set as output. We denote a decision on a random choice as $\CodeIn{random.choice(X)} \rightarrow \CodeIn{x}, \CodeIn{x} \in \CodeIn{X}$.


Operations within the trace depend on these decisions on random choices. In this paper, we denote all of decisions on random choices in the operation $O$ as $Dec(O)$.
\rev{
$$Dec(O): \CodeIn{random.choice(X)} \rightarrow \CodeIn{x}, \CodeIn{random.choice(Y)} \rightarrow \CodeIn{y}, ..., where\ decision\ X, Y, ... \in operation\ O$$
}
As an example, in the trace of the execution on Listing~\ref{lst:g0} (i.e., ~\Cref{eq:trace}), $O_4: \CodeIn{w += "\ca{}"}$, corresponding to a transition through a generator’s subpart $Prog(O_4): \CodeIn{w += random.choice(ascii\_lowercase)}$ (Line 9), relies on the decision made for ~\CodeIn{random.choice(ascii\_lowercase)}. For $O_4$, the decision is $Dec(O_{4}): \CodeIn{random.choice(ascii\_lowercase)} \rightarrow \CodeIn{"\ca{}"}$.



Following the goal of ~\emph{trace alignment} (as shown in ~\Cref{eq:opalign}), our approach manipulates the randomness during the re-execution to achieve alignment of operations in the re-execution trace and unlabeled operations in the reduced trace, i.e., $O^{\prime}_{i} \approx O_{x_i}$ \rev{($O_{i}^{\prime}$ denotes an operation in the re-execution trace $T^{\prime}$ and $O_{x_i}$ denotes the mapping operation in the reduced trace $T^{\dagger}$)}, by following two criteria: (1) matching $Prog(O^{\prime}_i)$ with $Prog(O_{x_i})$, and (2) matching $Dec(O^{\prime}_i)$ with $Dec(O_{x_i})$.

\textbf{Matching $Prog(O^{\prime}_i)$ with $Prog(O_{x_i})$.}
Based on the patterns of reducible parts, during re-execution, our approach manipulates the decision on the random choices in the operations which are the initiations of two reducible patterns.

For reducible selection pattern in the reduced trace such as 
\pre{
$T^{\dagger}: \langle ..., S_{init}, \mbox{\sout{$S_{block}$}}, ... \rangle$
}
\rev{
$T^{\dagger}: \langle ..., B_{init}, \mbox{\sout{$B_{block}$}}, ... \rangle$
}
,
our approach manipulates the decision on \CodeIn{random.choice([False, True])} from \CodeIn{True} in \pre{$S_{init}$}\rev{$B_{init}$} to \CodeIn{False} in 
\pre{$S_{init}^{\prime}$}\rev{$B_{init}^{\prime}$}, i.e., 
\pre{$Dec(S_{init}^{\prime}): \CodeIn{random.choice([False, True])} \rightarrow \CodeIn{False}$.}
\rev{
$Dec(B_{init}^{\prime}): \CodeIn{random.choice([False, True])} \rightarrow \CodeIn{False}$.
}
This manipulations enables the search on an execution with the trace \pre{$T^{\prime}: \langle ..., S_{init}^{\prime}, ... \rangle$}
\rev{$T^{\prime}: \langle ..., B_{init}^{\prime}, ... \rangle$}
.

For reducible loop pattern in the reduced trace such as
$
T^{\dagger}:
\langle ..., 
L_{init}, 
..., \mbox{\sout{$L_{x_1 - 1}$}}, L_{x_1}, 
...,
\mbox{\sout{$L_{x_2 - 1}$}}, 
L_{x_{2}},  \\
..., \mbox{\sout{$L_{x_k - 1}$}},
L_{x_{k}}, \mbox{\sout{$L_{x_{k} + 1}$}}, ...
\rangle.
$
\ttwo{} will manipulate the decision on the random choice of the number of loop iterations from $n$ to $k$ ($k$ is the number of elements which are not labeled in $L\{n\}$) in $L^{\prime}_{init}$, i.e., $Dec(L^{\prime}_{init}): \CodeIn{random.choice(range(N))} \rightarrow \CodeIn{k}$, to search on an execution with the trace 
$T^{\prime}: \langle ..., L^{\prime}_{init}, L^{\prime}_{1}, L^{\prime}_{2}, ..., L^{\prime}_{k}, ... \rangle.$



\textbf{Matching $Dec(O^{\prime}_i)$ with $Dec(O_{x_i})$.}
For other operations, i.e. operations that are not the initiation of reducible loop pattern (e.g. $L_{init}$) or the initiation of reducible selection pattern (e.g. \pre{$S_{init}$}\rev{$B_{init}$}),
when our approach aligns $O_{i}^{\prime}$ with $O_{x_i}$,
the decisions on random choices in $O_{i}^{\prime}$ are manipulated to match those in $O_{x_i}$, i.e., $Dec(O_{i}^{\prime}) = Dec(O_{x_{i}})$. To achieve this, \tz{} will instrument all of $Dec(O_{i})$ in the given execution. Then, during re-executions on the generator, ~\ttwo{} manipulates $Dec(O_{i}^{\prime})$ to match with the corresponding $Dec(O_{x_{i}})$.

\rev{Following the above design, the} trace of the re-execution on the generator by ~\ttwo{}, i.e., $T^{\prime}$, is expected to align with the corresponding reduced trace $T^{\dagger}$. For some simple generators such as our motivating example, it is always achievable.
However, due to the dependencies between operations in the trace, it is possible that the trace alignment
becomes unattainable through our approximate method, as we called ~\inft{}.
An ~\inft{} happens when our approach fails on align operations in trace $T^{\prime}$ with the corresponding operations in trace $T^{\dagger}$, i.e., $O_{i}^{\prime} \not\approx O_{x_i} $.
We will showcase the details and our strategies of relieving this problem in Section~\ref{s4.3}.

\section{Implementation}~\label{sec:s4}

In this section, we first illustrate a naive implementation of our approach for the motivating example (i.e., \Cref{lst:g0}) in Section~\ref{sec:s4.1}.
We then demonstrate the implementation of two strategies designed for ~\tone{} in Section~\ref{s4.2}, and three strategies we designed for ~\ttwo{} in Section~\ref{s4.3}.


\subsection{Naive Reduction Implementation for the Motivating Example}~\label{sec:s4.1}
A naive implementation of our approach for our motivating example is shown in Listing~\ref{lst:g2}.
The highlighted lines represent the additional parts introduced by our approach in the original generator (Listing~\ref{lst:g0}).

Specifically, Line 24 runs the instrumented \CodeIn{gen} for the first time and collects \CodeIn{random} states \CodeIn{S} before each loop iteration (Line 19).
Lines 31--35 implement \ttwo{} by enumerating all of subsequences of collected random states with \CodeIn{REPLAY} flag being set, and manipulating the random states in the re-executions of the generator.


\begin{lstlisting}[language=Python, caption=
A naive implementation of \toolname{} for the motivating example, label={lst:g2}, escapechar=|
]
import random
from string import ascii_lowercase
from more_itertools import powerset

|\colorbox{yellow}{REPLAY = False}| # This flag distinguishes the first execution and re-executions.
|\colorbox{yellow}{S = []}| # This variable stores the trace of the original execution.
|\colorbox{yellow}{RS = []}| # This variable stores the reduced trace derived from Trace Reduction.

def gen(seed=26524): # This default seed is for generating "abc\nabc\n".
    random.seed(seed)
    w = ""
    n = random.choice(range(20)) # Set the number of loop iterations.
    |\colorbox{yellow}{if REPLAY:}|
        |\colorbox{yellow}{n = len(RS)}| # Adjust the number of loop iterations.
    for i in range(n):
        |\colorbox{yellow}{if REPLAY:}|
            |\colorbox{yellow}{random.setstate(RS[i])}| # Manipulate decisions on random choices.
        |\colorbox{yellow}{else:}|
            |\colorbox{yellow}{S.append(random.getstate())}| # Save the states of random.
        w += random.choice(ascii_lowercase) # Append a letter into w.
    w += '\n'
    return w + w

gen() # Trace Instrumentation of the execution.

REPLAY = True 

def P(I): # We set an artificial property test function here.
    return (I[-1] == "\n") and (I[:len(I) // 2] == I[len(I) // 2:]) and ("c" in I)

for RS in powerset(S): # A naive implementation of Trace Reduction that enumerates all of subsequences of S as reduced traces in a brute-force way (it can be replaced by ddmin or HDD).
    I = gen() # I is a reduced test input obtained by Trace-aligned Re-execution.
    if P(I): # Check if I exhibits property.
        print(I) # It will output "c\nc\n".
        break # The enumeration is ordered by the size, thus we could exit the search early when we obtain a result.

\end{lstlisting}

Note, our actual implementation improves this native instrumentation in three key aspects:

\begin{enumerate}
    \item The actual implementation transparently decorates APIs that produce random choices (e.g. \CodeIn{random.choice} in above example). The decorated APIs are able to record or manipulate the random choices. 
    \item The actual implementation of \tone{} is optimized by the search strategies of \emph{ddmin} and HDD, instead of the brute-force enumeration in Line 31.
    \item The instrumented generator assumes there is no \inft{} (which does not happen in this case). Re-alignment is needed when \inft{} occurs.
\end{enumerate}

These implementation details are explained below.

\subsection{Implementation of Instrumentation}~\label{s4.1}
In actual implementation, our approach conducts instrumentation on the generator by replacing APIs that produce random choices (e.g. \CodeIn{random.choice} in above example) with decorated ones that can record and modify random choices.
The implementation cost depends on the number of kinds of APIs that produce random choices (instead of the generator's size or the number of random choices in the generator). In practice, this number is fairly small: among our studied generators (as described in Section~\ref{s5.1.3}), the number of kinds of APIs is 2 for graph generator, 2 for DL model generator, and 6 for JavaScript program generator.

Explicitly, the instrumentation records random choices at the first execution, then during re-executions, our approach modifies the random choices, i.e., changes return values of API calls related to randomness. When the execution on the generator invokes the APIs that produce random choices, the instrumentation fetches following execution information: arguments, return values, the execution location (program counter), and execution stack. The execution location and execution stack are used for ~\tone{}, and the arguments and return values are used for ~\ttwo{}.




\subsection{Strategies for \rawtone{}}~\label{s4.2}
Followed the description in Section~\ref{tone}, we now demonstrate our two strategies of optimizing the way of deriving reduced traces in ~\tone{}: \str{} and \ttr{}.
To derive reduced traces we are interested during reduction, we conduct search strategies from existing delta debugging algorithms by modeling reducible parts as either a sequential structure or a tree structure. 

For \textbf{\str{}}, the reducible parts are modeled as a sequential structure: every subpart of reducible parts is regarded as an element, and all of them arrange into a sequence with the order in the trace. Based on it, we could build the search on deriving reduced traces upon strategies used in the existing delta debugging algorithms such as ~\emph{ddmin}~\cite{DD}. The search strategy of ~\emph{ddmin} employs a binary search-like technique, iteratively dividing and testing subsets of the given sequence to efficiently find a smaller one.

\begin{figure}[]
    \includegraphics[width=\linewidth]{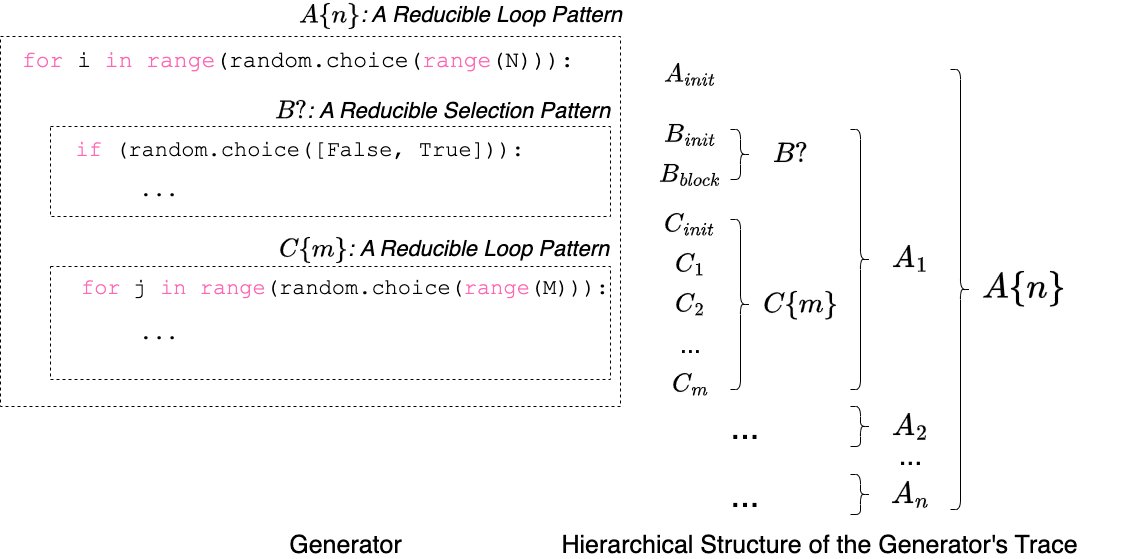}
    \caption{\rev{A generator and its trace structure under \textbf{\ttr{}}.}}
    \label{fig:treebased}
\end{figure}

For \textbf{\ttr{}}, the reducible parts are modeled as a hierarchical tree structure. 
Every subpart of the reducible parts is regarded as an element, and all of them arrange into a tree. As a quick example\pre{, consider the generator written in Python as follows:}\rev{ shown in Figure~\ref{fig:treebased}, the left side is a generator written in Python, as well as three reducible patterns within the generator denoted as $A\{n\}$, $B?$, and $C\{m\}$;
and the right side is the hierarchical structure of the generator's execution trace.
}

A tree structure for all of reducible parts in the trace is as follows: we set nodes to denote each reducible part $X$ as well as each subpart of $X$ (except for the initiation part, e.g., $A_{init}$, \pre{$B^{1}_{init}$}\rev{$B_{init}$}).
The subparts of $X$ are all children of $X$, e.g. $A_1$ is a child of $A\{n\}$. Also, we set the parent of the node denoted reducible part $X$ as the child of the node denoted a subpart that directly contains $X$, e.g. \pre{$C^{1}\{m_1\}$}\rev{$C\{m\}$} is the child of $A_{1}$.
To construct such structure, during the execution, we maintain a stack to keep track of execution point. When the execution point reaches \pre{to}\rev{} the initiation of a reducible pattern $X$, we set a nested structure for the following executions, continuing until the execution point exits $X$.
This implementation of constructing a hierarchical structure of the execution's trace is inspired by a technique called program execution indexing~\cite{DBLP:conf/pldi/XinSZ08}. 

Based on the tree structure, the search strategy of 
Hierarchical Delta Debugging (HDD)~\cite{HDD} is applicable. 
HDD extends ~\emph{ddmin} by performing ~\emph{ddmin} to the nested structures level by level, aiming to isolate relevant parts more progressively. 

\subsection{Strategies of \rawttwo{}}~\label{s4.3}
In this subsection, followed the description in Section~\ref{ttwo}, we first illustrate two kinds of cases of ~\inft{} in Section~\ref{s4.4.1}. To address the issues caused by ~\inft{}, we design three strategies in our approach as demonstrated in Section~\ref{s4.4.2}.

\subsubsection{Infeasible trace alignment}~\label{s4.4.1}
In order to achieve ~\emph{trace alignment} (i.e., $T^{\prime} \approx T^{\dagger}$), our approach aligns operations in trace $T^{\prime}$ with the corresponding operations in trace $T^{\dagger}$ (i.e., $O_{i}^{\prime} \approx O_{x_i} $). ~\iinft{} happens when the alignment can not be achievable.
Next, we demonstrate two kinds of cases when ~\inft{} happens.


\textbf{Case 1: failure of matching $Prog(O_{i}^{\prime})$ with $Prog(O_{x_i})$.}~\label{case1}
\iinft{} occurs due to a mismatch between the generator's subparts to which operations correspond, i.e.,
$
Prog(O_{i}^{\prime}) \neq Prog(O_{x_i})
\Longrightarrow
O_{i}^{\prime} \not\approx O_{x_i}.
$
To illustrate, let us consider a generator written in Python as follows:
\begin{lstlisting}[language=Python, label={lst:inf1}]
    x = 0
    if random.choice([False, True]):
        x = 1
    if x == 0:
        f()
    else:
        g()
    ...
\end{lstlisting}

An execution $E$ of this generator with the trace $T=\langle O_1, O_2, O_3, O_4, O_5 \rangle$ as follows:

$$\{\} \stackrel{O_{1}: \CodeIn{x = 0}} \longrightarrow
\{x=0\} \stackrel{O_{2}: \CodeIn{if True }} \longrightarrow
\{x=0\} \stackrel{O_{3}: \CodeIn{x = 1}} \longrightarrow
\{x=1\} \stackrel{O_{4}: \CodeIn{if x == 0 }} \longrightarrow
\{x=1\} \stackrel{O_{5}: \CodeIn{call g()}} \longrightarrow
...
$$

Assuming a \rtt derived from $T$ as
$T^{\dagger}: \langle O_{1}: \CodeIn{x = 0}, O_2: \CodeIn{if True}, \mbox{\sout{$O_{3}: \CodeIn{x = 1}$}}, O_4: \CodeIn{if x == 0}, O_{5}: \CodeIn{call g()}), ... \rangle $,
\ttwo{} aims to find an execution with trace
$T^{\prime}: \langle O^{\prime}_1, O^{\prime}_2, O^{\prime}_3, O^{\prime}_4, ... \rangle \approx T^{\dagger}$, i.e., $O^{\prime}_1 \approx O_1, O^{\prime}_2 \approx O_2, O^{\prime}_3 \approx O_4, O^{\prime}_4 \approx O_5$
. However, achieving such trace alignment is infeasible. This is due to the fact that, after manipulating decision on \CodeIn{random.choice([False, True])} from \CodeIn{True} in $O_2^{\prime}$ to \CodeIn{False} in $O_2^{\prime}$, the value of $x$ keeps as $0$. According to $O_3^{\prime}: \CodeIn{if x == 0}$, the execution on the generator would call $\CodeIn{f()}$ instead of $\CodeIn{g()}$, i.e., 
$ (Prog(O_{4}^{\prime}): \CodeIn{f()}) \neq (Prog(O_{5}): \CodeIn{g()})
\Longrightarrow O_{4}^{\prime} \not\approx O_{5}.$

\textbf{Case 2: failure of matching $Dec(O_{i}^{\prime})$ with $Dec(O_{x_i})$.}~\label{case2}
\iinft{} occurs due to a mismatch between the decisions on random choices, i.e., 
$
Prog(O_{i}^{\prime}) = Prog(O_{x_i}),
Dec(O_{i}^{\prime}) \neq Dec(O_{x_i})
    \Longrightarrow O_{i}^{\prime} \not\approx O_{x_i}.
$
As an example, consider a generator written in Python as follows:
\begin{lstlisting}[language=Python, label={lst:inf2}]
    x = [0] # x is initialized as a list with a single element.
    if random.choice([False, True]):
        x += [1] # list x appends a single element.
    if random.choice([False, True]):
        y = random.choice(x) # variable y is randomly choosed from x.
    ...
\end{lstlisting}

Given execution trace $T$ in which both random choices are \CodeIn{True} and a \rtt which is derived from  $T$ as ($\overline{y}$ could be \CodeIn{0} or \CodeIn{1}):
\begin{equation}~\label{eq:tdag}
    T^{\dagger}: \langle O_1: \CodeIn{x = [0]}, O_2: \CodeIn{if True}, \mbox{\sout{$O_3: \CodeIn{x += [1]}$}}, O_4: \CodeIn{if True}, O_5: \CodeIn{y = } \overline{y}, ... \rangle.
\end{equation}

\ttwo{} aims to find an execution $E^{\prime}$ with trace $T^{\prime}: 
\langle O^{\prime}_1, O^{\prime}_2, O^{\prime}_3, O^{\prime}_4, ... \rangle \approx T^{\dagger}$,
$$
E^{\prime}: \{\} \stackrel{O_{1}^{\prime}: \CodeIn{x=[0]}}\longrightarrow
\{x=[0]\} \stackrel{O_{2}^{\prime}: \CodeIn{if False}}\longrightarrow
\{x=[0]\} \stackrel{O_{3}^{\prime}: \CodeIn{if True}}\longrightarrow
\{x=[0]\} \stackrel{O_{4}^{\prime} }\longrightarrow 
...
$$

For the alignment on $O_4^{\prime}$ and $O_5$, their generator's subparts are matched, i.e., $Prog(O_4^{\prime}) = Prog(O_5): \CodeIn{y = random.choice(x)}$, however, their decisions on random choices, i.e., $Dec(O_{4}^{\prime})$ and $Dec(O_5)$, may be mismatched.
There are two possibilities of $O_5$: one is $\overline{y} = \CodeIn{0}$, when $O_4^{\prime}$ is aligned with $O_5$; another is $\overline{y} = \CodeIn{1}$, when an infeasible trace alignment happens.

As the first situation, $\overline{y} = \CodeIn{0}$, i.e., $Dec(O_5): \CodeIn{random.choice(x)} (\CodeIn{x = [0,1]}) \rightarrow \CodeIn{0}$, because we can still manipulate $Dec(O_{4}^{\prime})$ as $\CodeIn{random.choice([0])} \rightarrow \CodeIn{0}$, to match with $Dec(O_5)$. Note that for two random choices corresponding to the same subpart of the generator, the range of random choices might be different because the inputs of these random choices depend on values in program states. However, if the outputs of these random choices can be the same, we still regard these two random choices as matched. 


However, as the other situation, $\overline{y} = \CodeIn{1}$, i.e., $Dec(O_5): \CodeIn{random.choice(x)} (\CodeIn{x = [0,1]}) \rightarrow \CodeIn{1}$, we observe that achiving alignment between $O_{4}^{\prime}$ and $O_5$ is infeasible.
As the above execution $E^{\prime}$ shown, before executing $Prog(O_4^{\prime})$ (i.e. Line 5), the value of \CodeIn{x} is \CodeIn{[0]}, resulting that $Dec(O_4^{\prime})$ can only be $\CodeIn{random.choice([0])} \rightarrow \CodeIn{0}$. Thus, $O^{\prime}_{4}: \CodeIn{y = 0}$ does not match with $O_5$.
\begin{equation*}
    \begin{aligned}
    \begin{cases}
    Dec(O_{4}^{\prime}): \CodeIn{random.choice([0])} \rightarrow \CodeIn{0}, \\
    Dec(O_5): \CodeIn{random.choice([0, 1])} \rightarrow \CodeIn{1}
    \end{cases}
\Longrightarrow
    \begin{cases}
    O_{4}^{\prime}: \CodeIn{y = 0}\\
    O_5: \CodeIn{y = 1}
    \end{cases}
\Longrightarrow
O_{4}^{\prime}  \not\approx O_5
    \end{aligned}
\end{equation*}




\subsubsection{Relieving infeasible trace alignment}~\label{s4.4.2}
In the context of achieving trace alignment in ~\ttwo{}, when ~\inft{} happens (i.e., $O_{i}^{\prime} \not\approx O_{x_i} $), it means that we cannot fully preserve the semantics of remaining part after removing subparts of the given execution. To relieve this issue,
we have designed three strategies named \szero{}, \sone{} and \stwo{}, which serve as workarounds for situations where ~\inft{} arises.

In general, the first strategy, \szero{}, directly ``halts'' the current execution to filter out reduced traces that are unable to align with. The second strategy, \sone{}, aims to add extra ``removals'' in the reduced trace to skip the operations that can not be aligned with. Both of these two strategies work in a conservative way.
The third strategy, \stwo{}, however, allows the execution to continue by re-aligning the trace of the remaining execution with the given reduced trace.

In our implementation, we apply these three strategies in a singular mode: consistently applying a single strategy when ~\inft{} happens. We then demonstrate these three strategies in detail.


\begin{figure}[]
    \includegraphics[width=\linewidth]{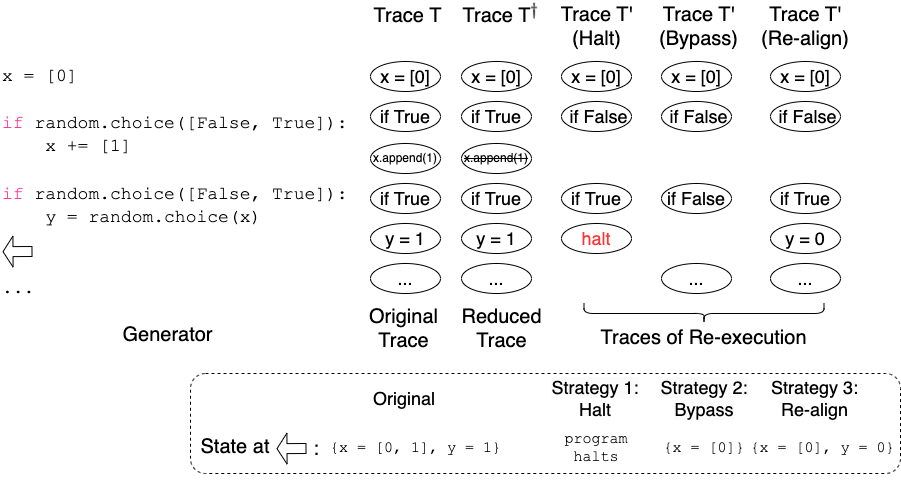}
    \caption{An example of ~\emph{infeasible trace alignment} and how different strategies work on it.}
    \label{fig:strexp}
    \vspace{-0.7cm}
\end{figure}

\textbf{Strategy 1: \szero{}.}
The first strategy, \szero{}, when ~\inft{} happens, directly conduct a complete stop on the execution on the generator. It is a straight-forward strategy, and the most conservative way for ~\emph{trace alignment}. With the strategy \szero{}, when an ~\emph{infeasible trace alignment} happens, it is regarded as a failure of \ttwo{} to find an execution which yields a reduced test input. As a result, \toolname{} will back to the iterative search on \tone{} for deriving other reduced traces.



\textbf{Strategy 2: \sone{}.}
The second strategy, \sone{}, when ~\inft{} happens ($O_{i}^{\prime} \not\approx O_{x_i} $), adds extra ``removals'' on reducible part where $O_{i}^{\prime}$ in it to bypass the supposed alignment.
Specifically, if $O_{x_i}$ is in a subpart $P_{i}$ of a reducible part $P$, the strategy \sone{} add an extra ``removal'' on $P_{i}$ to remove $O_{x_i}$ in the reduced trace $T^{\dagger}$. If there exist multiple such reducible parts, it chooses the smallest one, i.e., the one directly contains $O_{x_i}$. If no such reducible part exists, it will raise an error and stop the execution like \szero{}. For example, in ~\Cref{eq:treetrace}, if infeasible trace alignment happens on an operation inside of $C^{1}_{2}$, \toolname{} chooses to remove $C^{1}_{2}$ instead of $A_1$.
As a result, $O_{i}^{\prime}$ will not correspond to the program's subpart that $O_{x_i}$ interprets (i.e., $Dec(O_{x_i})$), instead, $O_{i}^{\prime}$ will match with the next unlabeled operations in the reduced trace $T^{\dagger}$ (with the extra ``removals'' on $P_{i}$).

For example, a trace with a reducible loop pattern is as follows: $T: \langle L_{init}: O_1, L_1: O_2, L_2: O_3, L_3: O_4 \rangle$, and a reduced trace $T^{\dagger}: \langle L_{init}, \mbox{\sout{$L_1$}}, L_2, L_3 \rangle = \langle O_1, \mbox{\sout{$O_2$}}, O_3, O_4 \rangle  $ is given for \ttwo{}.
Trace alignment requires $T^{\prime}: \langle O^{\prime}_{1}, O^{\prime}_{2}, ...\rangle \approx T^{\dagger}$, i.e., $O^{\prime}_{1} \approx O_1, O^{\prime}_{2} \approx O_3$, assume if $O^{\prime}_{2}$ can not be aligned with $O_3$, due to that $O_3$ is inside of $L_2$, then we labels ``removals'' on the $L_2$ (as well as adjusting random choices in $L_{init}$). As a result, $O^{\prime}_{2}$ will align with the next unlabeled operation, which is $O_4$ in $L_3$ in our example.


\textbf{Strategy 3: \stwo{}.}
The third strategy, \stwo{}, when ~\inft{} happens, continues to execute with re-aligning on following operations. 
In contrast to the other two strategies which conservatively achieves ~\emph{trace alignment}, this strategy operates in a non-conservative manner by permitting misalignment during the re-execution.
For each decision on random choices in misaligned operations, we will take an arbitrary return value according to the value of random choices' augments in the current program state.
And after the misaligned operations, \toolname{} with strategy \stwo{} attempts to align the following operations in $T^{\prime}$ with the corresponding operations in $T^{\dagger}$.

For example, for \textbf{Case 1} in Section~\ref{case1}, when the infeasible trace alignment happens on the operation $O^{\prime}_4$ in re-execution, i.e., $O^{\prime}_4: \CodeIn{call f()} \not\approx O_5: \CodeIn{call g()}$, with \stwo{} strategy, our approach will align the operation after finishing the call on \CodeIn{f()} with the corresponding one in reduced trace, i.e., aligning operations (the one in re-execution and the one in the initially given execution) that correspond to Line 8 in the Case 1 in Section~\ref{case1}.



\textbf{A detailed example.} To further demonstrate the difference of these three strategies of ~\ttwo{}, let us take a detail example (the same as the \textbf{Case 2} in Section~\ref{case2}).

As shown in Figure~\ref{fig:strexp}, the trace $T$ demonstrates one of its original execution's \pre{trace}\rev{traces}, a reduced trace is shown as trace $T^{\dagger}$ (the same as Equation~\ref{eq:tdag}).
Note, when the re-execution runs at \CodeIn{y = random.choice(x)}, the value of variable \CodeIn{x} is \CodeIn{[0]}, thus an infeasible trace alignment happens, i.e., the decision on \CodeIn{random.choice(x)} (i.e., \CodeIn{random.choice([0])}) can not match with the one (which returns \CodeIn{1}) in the original execution.

When the infeasible trace alignment happens, different strategies work as follows:
the \textbf{halt} strategy directly halt the re-execution of the generator;
the \textbf{bypass} strategy changes the choice on if-condition from \CodeIn{True} to \CodeIn{False} (i.e., bypasses the random choice), and continue the re-execution;
the \textbf{re-align} strategy will interpret \CodeIn{y = random.choice(x)} as \CodeIn{y = 0} by manipulating the return value of \CodeIn{random.choice(x)} (return a random one in \CodeIn{x}), and continue the re-execution.

\section{Evaluation}\label{sec:s5}

To evaluate the effectiveness and efficiency as well as the versatility of \toolname{}, we study the following three research questions with an evaluation \pre{over three real-world application domains}\rev{on five benchmarks across distinct domains}:

RQ1 (\textbf{Effectiveness}): How effective is \toolname{} at reducing test inputs in different domains, compared to state-of-the-art reducers?

RQ2 (\textbf{Efficiency}): How efficient is \toolname{} at reducing test inputs in different domains, compared to state-of-the-art reducers?

RQ3 (\textbf{Ablation Study}): How do different strategies affect the performance of \toolname{}?


The reproducible evaluation is available on ~\url{https://github.com/GReduce/GReduce}.



\subsection{Experiment Settings}

\subsubsection{Application Domains}
Our evaluation includes three distinct application domains where domain-specific generators incur internal data dependencies and syntax-based reducers are less efficient in test input reduction: \textbf{graphs}, \textbf{DL models} and \textbf{JavaScript programs}.
Generator-based testing is the most popular practice for testing softwares in these domains~\cite{padhye2019semantic, DBLP:journals/corr/abs-2302-00842, DBLP:conf/icse/GuLZ022, holler2012fuzzing}.
For each application domain, we pick a popular software under test as our study subject: graph-library JGraphT~\cite{10.1145/3381449} for graphs, a popular DL compiler named TVM~\cite{DBLP:conf/osdi/ChenMJZYSCWHCGK18} for DL models, and Closure compiler~\cite{Closure} for JavaScript programs. All of those softwares are actively maintained by developers.

\rev{
We also employ two existing benchmarks from Hypothesis~\cite{DBLP:conf/ecoop/MaciverD19}, i.e., SymPy benchmark~\cite{DBLP:conf/ecoop/MaciverD19} and SmartCheck benchmark~\cite{DBLP:conf/ecoop/MaciverD19, pike2014smartcheck}, as two additional application domains in our evaluation, mainly for comparing the performance of \toolname{} and Hypothesis~\cite{DBLP:conf/ecoop/MaciverD19} (a choice-sequence-based reduction approach), as well as other two reducers QuickCheck~\cite{DBLP:conf/icfp/ClaessenH00} and SmartCheck~\cite{pike2014smartcheck}.

\textbf{SymPy benchmark} is an existing benchmark used in the evaluation of Hypothesis~\cite{DBLP:conf/ecoop/MaciverD19}.
SymPy is a symbolic algebra library for Python.
Previous research work builds a test generator based on TSTL framework~\cite{TSTL} to construct test cases for triggering bugs in SymPy. The reducer in Hypothesis will conduct ``internal reduction'' on the test cases to find a reduced test case that triggers the same bug.

\textbf{SmartCheck benchmark}~\cite{DBLP:conf/ecoop/MaciverD19, pike2014smartcheck} is a set of five synthetic benchmarks, named “bound5”, “binheap”, “calculator”, “parser”, and “reverse”.
Each of these benchmarks consists of generators for generating specific data types as well as the code that takes corresponding data types as inputs and has a known bug in it.
}




\subsubsection{Compared Reducers}
For the first two application domains, i.e., graphs and DL models, to the best of our knowledge, there is no existing automatic reduction tool which specifically targets at these domains. Thus we adopt a widely-used state-of-the-art syntax-based test input reducer, \textbf{Perses}~\cite{Perses}, as the representative of existing practice.
Perses leverages ANTLR grammar~\cite{ANTLR} for the test inputs (to derive syntax-based reduction rules). However, there is no existing ANTLR grammars for graphs and DL models in ANTLR official repository~\cite{ANTLR_repo}.
In order to conduct Perses on these two application domains, we use API calls in Python language to proxy the representation of inputs. For graphs, we use two APIs in JGraphT (for creating nodes and edges); for DL models, we use APIs in ONNX~\cite{ONNX}.
We also include another state-of-the-art reducer \textbf{T-PDD}~\cite{wang2023probabilistic}, an extension of Perses that incorporates optimizations based on a probabilistic model, as an additional baseline.
\rev{Besides, we also conduct experiments for directly applying ~\emph{ddmin} algorithm to the sequences of API calls, i.e., taking each API as a single element in the sequence for delta debugging, denoted as \textbf{~\emph{ddmin}}.
}

For the third application domain, JavaScript programs, we compare \toolname{} with Perses~\cite{Perses}, T-PDD~\cite{wang2023probabilistic}, and another state-of-the-art language-specific reducer \textbf{JSDelta}~\cite{JSDelta}.
JSDelta, as a language-specific reducer, is a mature tool for reduction on JavaScript programs, based on the WALA static analysis infrastructure~\cite{WALA}. 

\rev{
Besides,
to compare \toolname{} with reduction approaches that utilize generators, we include a state-of-the-art reduction method, \textbf{Hypothesis}~\cite{DBLP:conf/ecoop/MaciverD19}, as another compared reducer in our evaluation.
The approach used in Hypothesis is also named ``internal test-case reduction''~\cite{DBLP:conf/ecoop/MaciverD19}, which is based on directly reducing the random choices sequence of generators according to the shortlex order.
We compare \toolname{} with Hypothesis on two previous benchmarks, i.e., SymPy benchmark and SmartCheck benchmark, which come from the paper of Hypothesis~\cite{DBLP:conf/ecoop/MaciverD19}.
In addition, for the SmartCheck benchmark, we also compare GReduce with \textbf{QuickCheck}~\cite{DBLP:conf/icfp/ClaessenH00} and \textbf{SmartCheck}~\cite{pike2014smartcheck}, which are two generic reducers for algebraic data types defined in Haskell language. 
}

\subsubsection{Test Input Generators}~\label{s5.1.3}
For \pre{each application domain}\rev{graphs, DL models and JavaScript programs}, we adopt \toolname{} on representative generators that have been proven to effectively reveal many real-world bugs~\cite{padhye2019semantic,DBLP:journals/corr/abs-2302-00842} and have been widely used in the research community~\cite{reddy2020quickly, nguyen2022bedivfuzz, kukucka2022confetti}.

\textbf{Graph Generator} is a generator implemented in Zest~\cite{padhye2019semantic}. It utilizes the JGraphT library APIs~\cite{10.1145/3381449} to generate random graphs. The generated graphs could be used as inputs for testing various graph algorithms in JGraphT such as shortest path, biconnectivity.
This generator works in two main steps: first, creating nodes in the graph named; then creating edges between nodes.
Note, when generator creates an edge that connects two nodes $x$ and $y$ in the graph, there are dependencies between the edge and the creation of $x$, as well the creation of $y$.
\textbf{DL Model Generator}, named ~\emph{Isra}~\cite{DBLP:journals/corr/abs-2302-00842}, is a generator towards randomly generating valid deep learning models for testing deep learning compilers such as TVM~\cite{DBLP:conf/osdi/ChenMJZYSCWHCGK18}.
By taking DL models as computation graphs, ~\emph{Isra} incrementally builds computation graphs with appending operations one by one into the computation graph. 
~\emph{Isra} incorporates specification of DL model into the generation to avoid yielding invalid DL models.
The DL models generated by ~\emph{Isra} could guarantee compliance with specification of DL models, including the acyclicity of the generated graph and constraints of various DL operations such as \CodeIn{Gemm}\footnote{\url{https://onnx.ai/onnx/operators/onnx\_\_Gemm.html\#inputs}} and \CodeIn{Split}\footnote{\url{https://onnx.ai/onnx/operators/onnx\_\_Split.html\#inputs}}.

\textbf{\gJS{}} is a generator implemented in Zest~\cite{padhye2019semantic}, producing JavaScript programs by mimicing the syntax of JavaScript. It follows QuickCheck~\cite{DBLP:conf/icfp/ClaessenH00} style where a structured input is created using recursive subroutines.
There exists dependencies between subroutines in \gJS{}: for example, it maintains a global data structure to re-use previously generated identifiers for prioritizing the generated programs generate the the same identifier multiple times.
Note, this generator, used as a fuzzer in Zest, still generates a high percent of semantically invalid programs (53.2\% in our experiment), because it does not incorporate all of JavaScript program specification. 

We list the size of these generators as follows (dependent files and libraries are not included): Graph Generator\footnote{\url{https://github.com/rohanpadhye/JQF/blob/master/examples/src/main/java/edu/berkeley/cs/jqf/examples/jgrapht/FastGnmRandomGraphGenerator.java}}: 127 lines (114 LOC), 4.2 KB; DL Model Generator\footnote{\url{https://github.com/israProj/isra/blob/main/g.py}}: 630 lines (544 LOC), 17.9 KB; JavaScript Program Generator\footnote{\url{https://github.com/rohanpadhye/JQF/blob/master/examples/src/main/java/edu/berkeley/cs/jqf/examples/js/JavaScriptCodeGenerator.java}}: 333 lines (285 LOC), 12.2 KB. More details of our studied generators are shown in our project website~\cite{projectwebsite}.

\rev{
For the SymPy benchmark and the SmartCheck benchmark, we adopt \toolname{} on the existing generators in these benchmarks. We follow the same settings as the experiments in Hypothesis~\cite{DBLP:conf/ecoop/MaciverD19} to ensure a fair comparison in our evaluation.
}
\subsubsection{Bug Cases for Reduction}
Our subjects consist of 40 bug-inducing test inputs (10 on graphs, 20 on DL models, and 10 on JavaScript programs)\footnote{This number is equal to or larger than all recent publications on delta debugging at top venues as far as we are aware~\cite{ProbDD}.}. These subjects are collected by conducting generator-based testing with above generators to trigger bugs on the software in the corresponding domain:
for graphs, we manually collect 10 non-duplicate bug reports in the JGraphT issue tracker\footnote{\url{https://github.com/jgrapht/jgrapht/issues}} and reproduce them by the Graph Generator;
for DL models, following the settings in ~\cite{DBLP:journals/corr/abs-2302-00842}, we run the DL model generator for testing TVM~\cite{DBLP:conf/osdi/ChenMJZYSCWHCGK18} and collect 20 bug-inducing test inputs, ensuring that there are \pre{noduplications}\rev{no duplications} among them;
for JavaScript programs,
based on the JavaScript program generator in Zest~\cite{padhye2019semantic},
we collect 10 bug-inducing test inputs for Closure compiler~\cite{Closure}, ensuring that there are no duplications among them.



To set property test functions (i.e., test oracles) for these bug cases, we category them as two types: crash bugs and non-crash bugs.
For crash bugs (or unexpected exceptions), we find smaller test inputs that reproduce the exactly the same or fairly close\footnote{The similarity of error messages is computed by a string comparator named SequenceMatcher in difflib~\cite{difflib}. We set the threshold as 0.8 in our evaluation.} error message. This is a common practice when a user triggers an unknown error and manages to reduce the test input before reporting it to developers. In our subjects, 37 of 40 belong to this type.
For non-crash bugs that the software terminates normally, we use the differential testing strategy for our property test function, i.e., running reduced test input on both buggy version and fixed version, and comparing their consistencies. We notice that such property test function is non-existent until the bug has been fixed by developers. We claim that reduction is still necessary for developers because the test that trigger a specific bug will usually be added into test suites to improve the quality of test suites.
In such scenarios, a test input after reduction is more readable, maintainable and resource-saving than the one before reduction. In our subjects, 3 of 40 belong to this type.

The original size of our subjects ($Size_{o}$), on average, is (1) 92 nodes, 383 edges for graphs, (2) 84.25 nodes and 160 edges for DL models, (3) 272.2 non-blank characters for JavaScript programs.
For JavaScript programs, due to that the JavaScript program generator does not guarantee semantic validity, larger programs constructed by the generator are more likely to be rejected by the compiler. This results in the subjects on JavaScript program being relatively small in size. Our approach does not have inherent limitations on the size of bug-inducing test inputs.

\rev{
For bug cases of the SymPy benchmark and the SmartCheck benchmark, we follow the same settings as the experiments in Hypothesis~\cite{DBLP:conf/ecoop/MaciverD19} to ensure a fair comparison in our evaluation.
}

More details of our bug cases for reduction are shown in our project website~\cite{projectwebsite}.

\subsubsection{Metrics}
Following previous work ~\cite{HDD, Perses,10.1145/3586049, ProbDD}, we use following metrics to measure the performance of approaches in the study.

The \textbf{size of reduction results} ($Size$), as well as the \textbf{reduction quality} (\redq), are critical metrics for investigating effectiveness of test input reduction.
We define the size for graphs and DL models~\footnote{A DL model corresponds to a computation graph with nodes as operations and edges as computation flows.} as a tuple with number of nodes and number of edges in the graph, i.e., (\# of nodes, \# of edges).
We define the size of JavaScript programs as the number of non-blank characters in the program. 
Reduction quality is the percentage of the size of reduction results to the original size. For graphs and DL models, it is calculated by
dividing the sum of number of nodes and edges in the reduction results by the sum of number of nodes and edges in the original graph.

The \textbf{reduction time} ($Time$) denotes the overall time for reduction, we take it as the main metric for investigating efficiency of test input reduction.
The \textbf{number of property tests} ($\#tests$) is the number of checking the property during reduction. In practice, if the cost of property tests is heavy, then the number of running property tests during reduction is also critical for overall efficiency.
The \textbf{reduction speed} ($Speed$) is the number of units of size that a reducer can reduce per second on average.

Our evaluation is conducted on a computer with Intel® Core™ i5 CPU @ 1.4GHz and memory of 16GB. In our experiments, every reducer runs with a single thread and the reduction process of every subject has a timeout limit of 1 hour.


\subsubsection{Settings of Different Strategies}
We do an exhaustive comparison by setting \toolname{} with combinations of strategies we designed:
(1) for two strategies in \tone{}, we denote \toolname{} with \str{} as $G^{\da}_{*}$, and \toolname{} with \ttr{} as $G^{\db}_{*}$,
and we conduct the \emph{ddmin}~\cite{DD} search strategy for the former and HDD~\cite{HDD} search strategy for the latter; (2)
for three strategies of relieving ~\inft{} in \ttwo{}: \szero{}, \sone{} and \stwo{}, we denote \toolname{} with the strategy
\szero{} as $G^{*}_{\ssa}$, with the strategy \sone{} as $G^{*}_{\ssb}$, and with the strategy \stwo{} as $G^{*}_{\ssc}$.




\begin{table}[]
\caption{The evaluation results of \toolname{} with different settings on graphs, DL models, and JavaScript programs.}
\label{tab:res}

\resizebox{\textwidth}{!}{%
\begin{tabular}{cc|rrrrrr
}
\hline
\multicolumn{1}{c|}{{\color[HTML]{333333} Domains}} & \multicolumn{1}{c|}{{\color[HTML]{333333} Metrics}} & \multicolumn{1}{c}{{\color[HTML]{333333} $G_{h}^{S}$}} & \multicolumn{1}{c}{{\color[HTML]{333333} $G_{b}^{S}$}} & \multicolumn{1}{c}{{\color[HTML]{333333} $G_{r}^{S}$}} & \multicolumn{1}{c}{{\color[HTML]{333333} $G_{h}^{T}$}} & \multicolumn{1}{c}{{\color[HTML]{333333} $G_{b}^{T}$}} & \multicolumn{1}{c}{{\color[HTML]{333333} $G_{r}^{T}$}}  \\ \hline
\multicolumn{1}{l|}{}                        & Size                                        & (17.0, 18.3)                                            & (13.9, 15.0)                                            & (9.4, 9.8)                                              & (17.0, 18.3)                                             & (6.2, 7.1)                                               & { \textbf{(5.5, 6.0)}}                                             \\
\multicolumn{1}{l|}{}                        & Time                                         & 530.7s                                                   & 283.0s                                                   & 237.7s                                                   & 501.3s                                                    & { \textbf{73.2s}}                                      & 81.6s                                                     \\
\multicolumn{1}{l|}{}                        & \#tests                                        & 422.8                                                   & 595.7                                                   & 405.0                                                   & 439.3                                                    & 120.1                                                    & { \textbf{119.2}}                                    \\
\multicolumn{1}{c|}{\multirow{-4}{*}{Graphs}} & Speed                                        & 8.2                                                     & 13.8                                                    & 11.8                                                    & 8.1                                                      & { \textbf{19.3}}                                      & 16.3                                                        \\ \hline
\multicolumn{1}{l|}{}                        & Size                                        & (80.0, 140.9)                                           & (9.6, 16.4)                                             & (6.9, 11.1)                                             & (80.0, 141.2)                                            & (9.4, 16.0)                                              & { \textbf{(6.2, 9.7)}}                                      \\
\multicolumn{1}{l|}{}                        & Time                                         & 3.1s                                                     & 24.3s                                                    & 23.8s                                                    & { \textbf{2.2s}}                                       & 23.3s                                                     & 18.4s                                                   \\
\multicolumn{1}{l|}{}                        & \#tests                                        & 21.9                                                    & 61.6                                                    & 38.1                                                    & { \textbf{12.7}}                                      & 51.4                                                     & 28.6                                                      \\
\multicolumn{1}{c|}{\multirow{-4}{*}{DL models}}    & Speed                                        & 8.3                                                     & 36.9                                                    & 51.2                                                    & 10.7                                                     & 37.2                                                     & { \textbf{58.7}}                                    \\ \hline
\multicolumn{1}{l|}{}                        & Size                                        & 57.5                                                    & 45.8                                                    & 21.8                                                    & 57.5                                                     & 47.8                                                     & { \textbf{21.4}}                                       \\
\multicolumn{1}{l|}{}                        & Time                                         & 79.8s                                                    & 91.2s                                                    & 63.6s                                                    & 92.1s                                                     & 36.3s                                                     & { \textbf{21.4s}}                                               \\
\multicolumn{1}{l|}{}                        & \#tests                                        & 79.5                                                    & 36.8                                                    & 37.5                                                    & 83.3                                                     & 25.4                                                     & { \textbf{25.1}}                                             \\
\multicolumn{1}{c|}{\multirow{-4}{*}{JavaScript programs}}    & Speed                                        & 7.8                                                     & 3.7                                                     & 4.5                                                     & 5.2                                                      & 12.3                                                     & { \textbf{15.9}}                                         \\ \hline

\end{tabular}%
}
\end{table}
\begin{table}[]
\caption{The evaluation results of \toolname{} ($G^{\db}_{\ssc}$) and compared works on graphs, DL models, and JavaScript programs (the grey columns are ratios of metrics between \toolname{} and Perses, \toolname{} and JSDelta, \rev{\toolname{} and ~\emph{ddmin}}, \toolname{} and T-PDD).
}
\label{tab:res2}

\resizebox{\textwidth}{!}{%
\begin{tabular}{cc|rr
>{\columncolor[HTML]{C0C0C0}}r r
>{\columncolor[HTML]{C0C0C0}}r r
>{\columncolor[HTML]{C0C0C0}}r r
>{\columncolor[HTML]{C0C0C0}}r
}
\hline
\multicolumn{1}{c|}{{\color[HTML]{333333} Domains}} & \multicolumn{1}{c|}{{\color[HTML]{333333} Metrics}} & \multicolumn{1}{c}{{\color[HTML]{333333} $G_{r}^{T}$}} & \multicolumn{1}{c}{{\color[HTML]{333333} $Perses$}} & \multicolumn{1}{c}{\cellcolor[HTML]{C0C0C0} Ratio } & \multicolumn{1}{c}{{\color[HTML]{333333} $JSDelta$}} & \multicolumn{1}{c}{\cellcolor[HTML]{C0C0C0} Ratio} & \multicolumn{1}{c}{{\color[HTML]{333333} \rev{$ddmin$}}} & \multicolumn{1}{c}{\cellcolor[HTML]{C0C0C0} \rev{Ratio}} & \multicolumn{1}{c}{{\color[HTML]{333333} \rev{$T-PDD$}}} & \multicolumn{1}{c}{\cellcolor[HTML]{C0C0C0} \rev{Ratio}}                        \\ \hline

\multicolumn{1}{l|}{}                        & Size & \textbf{(5.5, 6.0)} & (22.0, 18.3)                                        & 28.5\%                             & -                                                    & -  & \rev{(26.4, 16.3)} &  \rev{26.9\%}                                                     &      (37.5, 38.2)        &       15.2\%          \\
\multicolumn{1}{l|}{}                        & Time  &      \textbf{81.6s}                              & 465.6s                                               & 17.5\%                                       & -                                                    & -  & \rev{329.8s} & \rev{24.7\%}                                                      &     588.7s         &   13.9\%              \\
\multicolumn{1}{l|}{}                        & \#tests  &  \textbf{119.2}   & 2325.4                                              & 5.1\%                                        & -                                                    & -    &  \rev{2337.9}  & \rev{5.1\%}                                                  &    1904.5          &       6.3\%          \\
\multicolumn{1}{c|}{\multirow{-4}{*}{Graphs}} & Speed  &  \textbf{16.3}  & 2.3                                                 & 7.2x                                         & -                                                    & -  &  \rev{3.96} & \rev{4.1x}                                                      &     1.4         &  11.7x               \\ \hline
\multicolumn{1}{l|}{}                        & Size  &     \textbf{(6.2, 9.7)}                             & (17.0, 29.0)                                        & 34.6\%                             & -                                                    & -  & \rev{(21.4, 38.4)} &  \rev{26.6\%}                                                     &   (65.5,114.8)           &   8.8\%              \\
\multicolumn{1}{l|}{}                        & Time  &  \textbf{18.4s}       & 3303.6s                                              & 0.6\%                                        & -                                                    & -   & \rev{1158.1s} & \rev{1.6\%}                                                     &    3600.0s          &    0.5\%             \\
\multicolumn{1}{l|}{}                        & \#tests  &    \textbf{28.6}                                              & 7191.9                                              & 0.4\%                                        & -                                                    & - & \rev{5685.0} & \rev{0.5\%}                                                       &    5905.1          &   0.5\%              \\
\multicolumn{1}{c|}{\multirow{-4}{*}{DL models}}    & Speed  &   \textbf{58.7}                                  & 0.06                                                 & 949.3x                                       & -                                                    & -  & \rev{0.17} & \rev{349.4x}                                                       &      0.02        &     3307.4x            \\ \hline
\multicolumn{1}{l|}{}                        & Size  &   \textbf{21.4}       & 28.3                                                & 75.6\%                                       & 35.3                                                 & 60.6\%  & \rev{-} & \rev{-}                                                 &      25.5        &    83.9\%             \\
\multicolumn{1}{l|}{}                        & Time &   21.4s     & 32.6s                                                & 65.4\%                                       & \textbf{13.1s}                                        & 162.8\%  & \rev{-} & \rev{-}                                                 &     97.0s         &  22.0\%               \\
\multicolumn{1}{l|}{}                        & \#tests  &   25.1      & 53.7                                                & 46.7\%                                       & \textbf{21.4}                                        & 117.3\%  & \rev{-} & \rev{-}                                                 &     117.3         &  21.4\%               \\
\multicolumn{1}{c|}{\multirow{-4}{*}{JavaScript programs}}    & Speed  &   15.9   & 12.1                                                & 1.3x                                         & \textbf{25.4}                                        & 0.6x  & \rev{-} & \rev{-}                                                    &    3.2          &        5.1x         \\ \hline

\end{tabular}%
}
\end{table}

\subsection{Results and Analysis}


Table~\ref{tab:res} shows the
overall performance of \toolname{} with different settings in terms of the four metrics in three domains. We highlight (in bold)  the optimal results among different settings on \toolname{} in ~\Cref{tab:res}.
Table~\ref{tab:res2} shows the overall performance of \toolname{} ($G^{\db}_{\ssc}$) and four compared works, i.e., \pre{Perses and JSDelta}\rev{Perses, JSDelta, ~\emph{ddmin} and T-PDD}. We also add extra columns in Table~\ref{tab:res2} (marked as grey) to show the ratios of metrics between \toolname{} ($G^{\db}_{\ssc}$) and compared works. 

\begin{figure}[]
    \centering
    \resizebox{\textwidth}{!}{\includegraphics{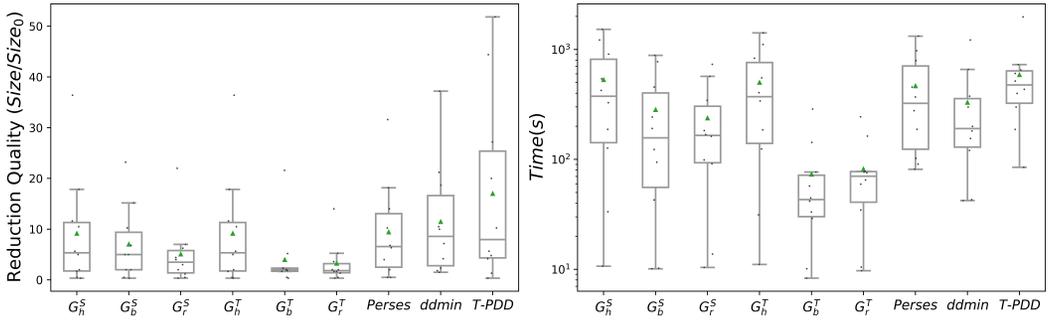}}  
    \caption{Results of the reduction quality (\redq) and reduction time ($Time$) on graphs.}
    \label{fig:graphres}
\end{figure}

\begin{figure}[]
    \centering
    \resizebox{\textwidth}{!}{\includegraphics{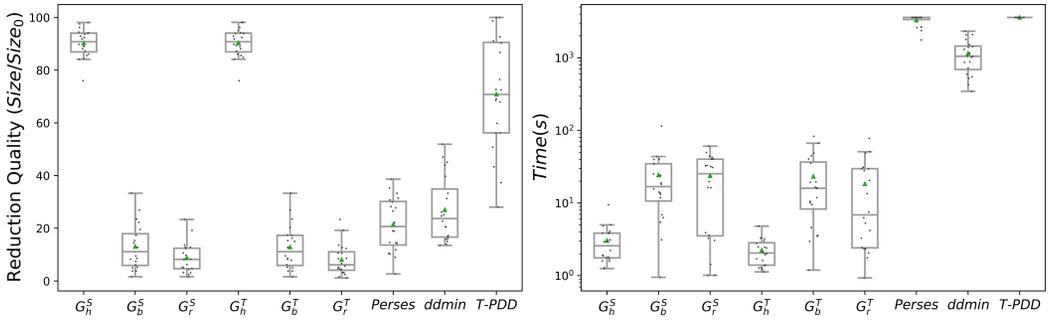}}
    \caption{Results of the reduction quality (\redq) and reduction time ($Time$) on DL models.}
    \label{fig:dlres}
\end{figure}

\begin{figure}[]
    \centering
    \resizebox{\textwidth}{!}{\includegraphics{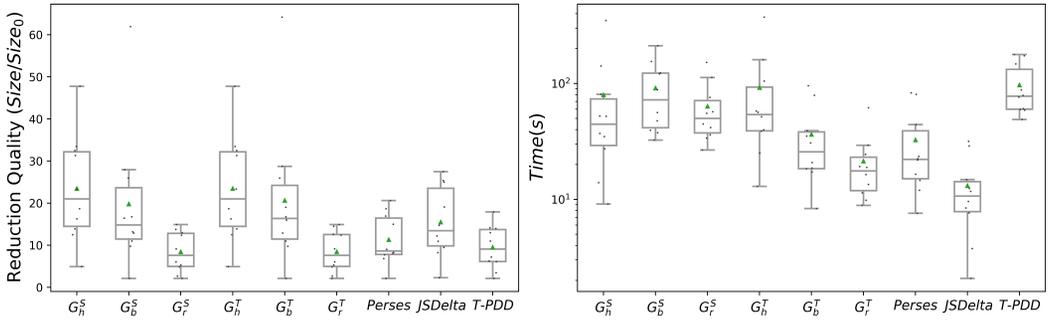}}
    \caption{Results of the reduction quality (\redq) and reduction time ($Time$) on JavaScript programs.}
    \label{fig:jsres}
\end{figure}

To evaluate the effectiveness, we mainly investigate the size of reduction results, as well as reduction quality (\redq{}).
To evaluate the efficiency, we mainly investigate the reduction time, as well as the number of property tests and reduction speed. The detailed statistical results on different domains are shown in Figure~\ref{fig:graphres} (graphs), Figure~\ref{fig:dlres} (DL models), and Figure~\ref{fig:jsres} (JavaScript programs).

Based on the evaluation result, among all of settings in \toolname{}, for effectiveness, $G^{\db}_{\ssc}$ constantly achieves the minimum size of reduced test input in all application domains, i.e., on average, (5.5, 6.0) on graphs, (6.2, 9.7) on DL models, 21.4 on JS programs. The reduction quality (\redq) is 2.4\%, 6.5\%, and 7.9\% respectively.
For efficiency, in general, $G^{\db}_{\ssc}$ and $G^{\db}_{\ssb}$ substantially outperform other settings on reduction time, number of property tests, and reduction speed (except when compared with $G^{*}_{\ssa}$ on DL models as we will discuss in Section~\ref{rq3}).

\subsubsection{Comparison with Perses, T-PDD, \rev{and ddmin}}



From Table~\ref{tab:res}, we can see that $G^{\db}_{\ssc}$, i.e., the setting on \toolname{} with optimal results, outperforms Perses, T-PDD \rev{and ~\emph{ddmin}} on all of metrics.
We also calculated the p-value of a paired sample Wilcoxon signed-ranked test to answer whether $G^{\db}_{\ssc}$ achieves significant improvement in both effectiveness and efficiency compared to Perses \rev{,~\emph{ddmin} and T-PDD,} and all p-values are significant (p < 0.05).
For graphs, on average, $G^{\db}_{\ssc}$ produces much smaller size of result than Perses (71.5\% smaller) \rev{,~\emph{ddmin} (73.1\% smaller), and T-PDD (86.1\% smaller)} with 82.5\% shorter \rev{, 75.3\% shorter and 84.8\% shorter} reduction time in the meanwhile.
For DL models, Perses reaches to the timeout (1 hour) on a majority of cases (14/20), and achieves the size of reduction results as (16.9, 29.0)\rev{ on average; ~\emph{ddmin} achieves the size of reduction results as (21.4, 38.4) with 1158.1s reduction time on average; T-PDD reaches to the timeout (1 hour) for all of cases, and achieves the size of reduction results as (65.5,114.8) on average.}

As a contrast, $G^{\db}_{\ssc}$ gets 65.4\% smaller size of reduction results, with only 18.4s reduction time (with the maximum as 77.7s).
For JavaScript programs, compared to Perses, $G^{T}_{\ssc}$ also achieves a 24.4\% smaller size of reduction results with 34.6\% shorter reduction time in the meanwhile. 
Note, for graphs and DL models, all of other settings on \toolname{} (except for $G^{*}_{\ssa}$ on DL models) also substantially outperforms Perses \rev{, ~\emph{ddmin}, and T-PDD}.

To further investigate the reason why Perses, T-PDD, \rev{and ~\emph{ddmin}} doesn't work well on graphs and DL models, we count the the proportion of test inputs derived by Perses, T-PDD \rev{and ~\emph{ddmin}} that comply with the specification.
\pre{The}\rev{For Perses, the} proportion is 12.57\% for graphs, and 2.55\% for DL models
\rev{; for ~\emph{ddmin}, the proportion is 17.99\% for graphs, and 42.62\% for DL models; for T-PDD, the proportion is 41.27\% for graphs, and 14.33\% for DL models}.
The result is consistent with our expectations: due to complex specification, a lot of invalid test inputs are derived from syntax-based reduction rules by Perses, T-PDD \rev{and ~\emph{ddmin}}, hurting the overall effectiveness and efficiency of reduction. Also, DL models, compared to graphs, hold more complex specification.
Hence, Perses and T-PDD yield lower ratios of valid inputs for DL models compared to graphs, resulting in its poorer performance.
\rev{For DL models, compared to Perses and T-PDD, ~\emph{ddmin} only focuses on reducing the sequence of API calls and omits the potential reduction opportunities on the parameter lists of API calls, thus, it achieves relatively higher ratios of valid inputs but worse size of reduction results due to its limited reduction search space.}
It indicates that preserving validity is critical for the performance of reduction on test inputs with specification. If the specification is relatively complex, exceeding mere syntactical structures, such as specification of DL models, syntax-based reduction would not be effective and efficient.

\subsubsection{Comparison with JSDelta}
Compared to JSDelta, $G^{\db}_{\ssc}$ gets 39.4\% smaller size of reduction results (the improvement is significant, i.e., p < 0.05) with 1.63x more reduction time on average.

Although the number of property tests for $G^{\db}_{\ssc}$ is close to JSDelta (with 1.17x difference), but the overall reduction time has 1.63x difference. JSDelta achieves faster reduction speed with less reduction time mainly due to that JSDelta utilizes a lightweight language-specific framework WALA~\cite{WALA} to customize the reduction and validate on JavaScript programs, which can swiftly produce valid JavaScript programs.

\rev{
\subsubsection{Comparison with Hypothesis, QuickCheck, and SmartCheck} \label{s5.2.3}
We compare \toolname{} with the state-of-the-art choice-sequence-based reduction approach, i.e., Hypothesis~\cite{DBLP:conf/ecoop/MaciverD19}, as well as QuickCheck~\cite{DBLP:conf/icfp/ClaessenH00} and SmartCheck~\cite{pike2014smartcheck}, on the two benchmarks from Hypothesis~\cite{DBLP:conf/ecoop/MaciverD19}.

On the SymPy benchmark, as shown in Table~\ref{fig:newcomp1}, \toolname{} significantly outperforms Hypothesis on the reduction time and the number of property tests: on average, \toolname{}($G^{T}_{b}$) takes 61.71s with 91.37 property tests; and \toolname{}($G^{T}_{r}$) takes 90.74s with 107.82 property tests; as a contrast, Hypothesis takes 976.22s with 1764.32 property tests.
However, \toolname{}($G^{T}_{r}$, i.e., our approach with the ``re-align'' strategy) achieves slightly worse results than Hypothesis on the size of reduction results (12.06 versus 11.53, 4.67\% difference). The reason is that
Hypothesis searches a larger space of random choice sequences, including those with smaller shortlex orders. This space is much larger than \toolname{}'s since \toolname{} only searches the reduced traces derived from the given trace, as shown in the significant differences in their reduction time and number of property tests. Additionally, in terms of reduction speed, which serves as a comprehensive metric balancing reduction time and the size of reduction results, \toolname{}($G^{T}_{r}$) also significantly outperforms Hypothesis (0.98 versus 0.08, 11.81x difference).

For a more comprehensive comparison, as shown in the rightmost column of Table~\ref{fig:newcomp1}, we further conduct the evaluation on another variant provided by Hypothesis, which only conducts delete operations on the choice sequences, named \textbf{Hypothesis(delete)}. Compared to Hypothesis, Hypothesis(delete) holds a smaller search space, resulting in larger size of reduction results (15.83 versus 11.53), less reduction time (227.84s versus 976.22s) and less number of property tests (316.19 versus 1764.32).
It shows that Hypothesis is sensitive to operation strategies on choice sequences.
The results also show that both \toolname{}($G^{T}_{b}$) and \toolname{}($G^{T}_{r}$) outperform Hypothesis(delete) on all four metrics, demonstrating that our reduction approach is a superior choice, achieving both effectiveness and efficiency.

On the SmartCheck benchmark, as shown in Table~\ref{fig:newcomp2}, our approach substantially outperforms all of the three compared reducers, i.e., Hypothesis, QuickCheck, and SmartCheck, on all metrics.
For the size of reduction results, \toolname{} achieves better or comparable results compared to these approaches, demonstrating the effectiveness of our approach.
The reason why the differences in the reduction results in some cases are not very obvious is that the reduction results of all compared approaches are close to the optimal solutions in those cases. For the rest of the metrics, i.e., the reduction time, the number of property tests, and the reduction speed, \toolname{} significantly outperforms compared approaches, demonstrating the efficiency of our approach.

}

\begin{table}[]

\caption{\rev{Evaluation Results on the SymPy Benchmark.}}\label{fig:newcomp1}
\rev{

\resizebox{\textwidth}{!}{%

\begin{tabular}{c|rrrrr
>{\columncolor[HTML]{C0C0C0}}r r
>{\columncolor[HTML]{C0C0C0}}r r >{\columncolor[HTML]{C0C0C0}}r}
\hline
  Metrics & \multicolumn{1}{l}{GReduce($G_{h}^{T}$)} & \multicolumn{1}{l}{GReduce($G_{b}^{T}$)} & \multicolumn{1}{l}{GReduce($G_{r}^{T}$)} & \multicolumn{1}{l}{Hypothesis} & \multicolumn{1}{l}{Hypothesis(delete)} 
  \\ \hline
Size    & 17.38                                        & 13.77                                              & 12.06                                 & \textbf{11.53}                 & 15.83                                  \\
Time    & 105.41s                                       & \textbf{61.71s}                                     & 90.74s                                 & 976.22s                         & 227.84s                                 \\
\#tests & 218.07                                       & \textbf{91.37}                                     & 107.82                                & 1764.32                        & 316.19                                 \\
Speed   & 0.61                                         & \textbf{1.48}                                      & 0.98                                  & 0.08                           & 0.31                                     

\\ \hline

\end{tabular}%
}
}
\end{table}

\begin{table}[]
\caption{\rev{Evaluation Results on the SmartCheck Benchmark\tablefootnote{
Some of numbers in the table are not obtained due to that our experiments on QuickCheck and SmartCheck rely on the artifact provided by Hypothesis~\cite{DBLP:conf/ecoop/MaciverD19}. The authors of Hypothesis have explained in their paper that they ran into some technical difficulties obtaining these numbers~\cite{DBLP:conf/ecoop/MaciverD19}.}.}}\label{fig:newcomp2}

\rev{
\begin{tabular}{c|c|rrrr}
 & Metrics &       \toolname{}($G^{T}_{\ssc}$)             & Hypothesis                 & QuickCheck                  & SmartCheck \\
\hline 

\multirow{4}{*}{binheap}    & Size    & 9.03                        & 9.01                           & \textbf{9.00}                  & 9.42                           \\
                            & Time    & \textbf{0.02}               & 2.68                           & 0.37                           & 0.17                           \\
                            & \#tests & \textbf{11.46}              & 182.97                         & 88.93                          & \multicolumn{1}{r}{-}          \\
                            & Speed   & \textbf{2.79}               & 1.33                           & \multicolumn{1}{r}{-}          & \multicolumn{1}{r}{-}                           \\
                        \hline
\multirow{4}{*}{bound5}     & Size    & \textbf{2.03}               & 2.08                           & 11.59                          & 5.96                           \\
                            & Time    & \textbf{0.06}               & 1.56                           & 7.17                           & 0.24                           \\
                            & \#tests & \textbf{9.49}               & 112.31                         & 1726.06                        & \multicolumn{1}{r}{-}          \\
                            & Speed   & \textbf{5.85}                        & 0.11                           & \multicolumn{1}{r}{-}          & \multicolumn{1}{r}{-}                  \\
                            \hline
\multirow{4}{*}{calculator} & Size    & 5.01                        & \textbf{5.00}                  & 5.09                           & \textbf{5.00}                  \\
                            & Time    & \textbf{0.01}               & 3.87                           & 4.27                           & 0.20                           \\
                            & \#tests & \textbf{8.80}               & 177.54                         & 31.16                          & \multicolumn{1}{r}{-}          \\
                            & Speed   & \textbf{6.20}               & 2.63                           & \multicolumn{1}{r}{-}          & \multicolumn{1}{r}{-}       \\
                            \hline
\multirow{4}{*}{parser}     & Size    & \textbf{3.03}               & 3.49                           & 3.99                           & 4.07                           \\
                            & Time    & \textbf{1.17}               & 87.92                          & 4.47                           & 2.76                           \\
                            & \#tests & \textbf{21.17}              & 131.69                         & 34.56                          & \multicolumn{1}{r}{-}          \\
                            & Speed   & \textbf{32.86}              & 0.92                           & \multicolumn{1}{r}{-}          & \multicolumn{1}{r}{-}                           \\
                            \hline
\multirow{4}{*}{reverse}    & Size    & \textbf{2.00}               & \textbf{2.00}                  & \textbf{2.00}                  & \textbf{2.00}                  \\
                            & Time    & \textbf{0.04}               & 0.69                           & 4.15                           & 0.21                           \\
                            & \#tests & \textbf{6.78}               & 63.04                          & 17.61                          & \multicolumn{1}{r}{-}          \\
                            & Speed   & \textbf{536.08}             & 1.65                           & \multicolumn{1}{r}{-}          & \multicolumn{1}{r}{-}      
\\ \hline

\end{tabular}
}

\end{table}

\subsubsection{Overhead of Instrumentation}

According to our evaluation, for test generation only (without running the software under test), compared to running generators without instrumentation, the overhead of GReduce introduced is around 36.5\% (Graph), 13.1\% (DL model), and 29.0\% (JavaScript program). Note, the time of test generation is much less than running property tests (i.e., running the software with test inputs), which is around 1 : 3.0 to 1 : 4.2. As a result, the extra overhead introduced by \toolname{} is fairly small in the overall reduction: around 2.5\% to 7.3\% in our evaluation.




\begin{tcolorbox}
\textbf{RQ1 and RQ2}: \toolname{} can reduce test inputs in different domains effectively and efficiently. 
On average, in \pre{three domains (graphs, DL models and JavaScript programs)}\rev{the domain of graphs, DL models and JavaScript programs}, the reduction results of \toolname{} are 28.5\%, 34.6\%, 75.6\% in the size of those from Perses and \toolname{} takes 17.5\%, 0.6\%, 65.4\% reduction time taken by Perses; the reduction results of \toolname{} are 15.2\%, 8.8\%, 83.9\% in the size of those from T-PDD and \toolname{} takes 13.9\%, 0.5\%, 22.0\% reduction time taken by T-PDD; also, the \toolname{}'s reduction results are 60.6\% in the size of those from JSDelta\rev{; additionally, on the SymPy benchmark and the SmartCheck benchmark, \toolname{} takes much less reduction time and number of property tests while achieving better or comparable reduction results compared to Hypothesis, QuickCheck, and SmartCheck}.
\end{tcolorbox}



\subsubsection{Impact of the Strategies in ~\rawtone{}}


To investigate the impact of strategies in ~\tone{}, i.e., ~\str{} and ~\ttr{}, we compare the results between $G^{\da}_{*}$ and $G^{\db}_{*}$ (with keeping other settings are identical).
Overall, $G^{\db}_{*}$  outperforms $G^{\da}_{*}$ on all of metrics among three domains in general (except for some of \pre{}\rev{the} cases that they are fairly close). For example, the result shows that $G^{\db}_{\ssc}$ outperforms $G^{\da}_{\ssc}$ with 40.0\%, 11.7\%, 1.8\% smaller size of reduction results
as well as 65.7\%, 22.7\%, 66.4\% shorter reduction time on average.

The reason is that \ttr{} with the HDD's search strategy utilizes the hierarchical structure of reducible parts in the trace for deriving reduced traces in \tone{}, thus, it is more effective and efficient for reduction, compared to \str{} with the ~\emph{ddmin}'s search strategy, as confirmed by our evaluation. 

\subsubsection{Impact of the Strategies in ~\rawttwo{}}~\label{rq3}
To investigate the impact of strategies in \ttwo{}, i.e., the strategies \szero{}, \sone{}, and \stwo{} , we compare the results between $G^{*}_{\ssa}$, $G^{*}_{\ssb}$, and $G^{*}_{\ssc}$ (with keeping other settings are identical).

For effectiveness, the strategy \stwo{} outperforms the strategy \sone{} while the strategy \sone{} outperforms the strategy \szero{} in all of the three domains: $G^{\db}_{\ssc}$ achieves 13.5\%, 37.4\%, 44.8\% smaller results than $G^{\db}_{\ssb}$ on graphs, DL models, and JavaScript programs respectively; also, $G^{\db}_{\ssb}$ achieves 62.3\%, 88.5\%, 16.9\% smaller results than $G^{\db}_{\ssa}$ on graphs, DL models, and JavaScript programs respectively.

For efficiency, the strategy \stwo{} and the strategy \sone{} achieves comparable results and the strategy \szero{} performs relatively worst among the three strategies in general.
As an exception, although $G^{*}_{\ssa}$ obtains the shortest time for reduction time on DL models, but the large size of reduction results (with reduction quality only 90\%) immensely diminishes its usefulness.

Based on the evaluation results, it shows that the strategy \szero{} has produced poorer results when compared to the other two strategies.
The reason is that dependencies are commonly presented in our generators, making it challenging to achieve ~\emph{trace alignment}. Consequently, $G^{*}_{\ssa}$ often halts prematurely due to ~\emph{infeasible trace alignment}, impeding the progress of reduction. The strategy \sone{} partially addresses this challenge by dynamically skipping some alignment in a conservative way. Thus, the strategy \sone{} achieves better results than the strategy \szero{} in general.
The strategy \stwo{}, compared to other two conservative strategies, allows misalignment of the re-execution on the generator to some extent when ~\emph{infeasible trace alignment} happens.
The evaluation result indicates that 
it is worthwhile to allow some misalignment during ~\ttwo{} on the generator because it can potentially yield smaller test inputs.





\begin{tcolorbox}
\textbf{RQ3}: 
\pre{Different strategies are critical for the performance of \toolname{}.}\rev{The performance of \toolname{} is affected by the strategies we proposed.} The evaluation result shows that (1) for ~\tone{}, tree-based strategy outperforms sequence-based strategy; (2) for ~\ttwo{}, the strategy \stwo{} generally yields superior results, followed by the strategy \sone{}, and lastly, the strategy \szero{}.
\end{tcolorbox}

\section{Threats to Validity}
There are several threats to the validity of our approach, which we outline along with proposals to mitigate them.
The threat to internal validity mainly lies in the correctness of the implementation of our approach, settings for property test functions, and the experimental scripts.
To reduce these threats, we have carefully reviewed our code.
The threats to external validity primarily include the degree to which the subject application domains as well as generators and bug-inducing test inputs are representative of practice. These threats could be reduced by more experiments on wider types of generators and test inputs in future work.



\section{Discussion and Future Work}~\label{discussion}
\rev{\subsection{Assumptions on the Properties of Generators}}
The effectiveness and efficiency of our approach relies on good properties of the generators. In the following, we outline two properties as a brief discussion.




\textbf{Monotonicity.}
$
E^{\prime} \preceq E
\Longrightarrow
I^{\prime} \preceq I.
$
Our approach requires monotonicity of the generator's execution, i.e., reduction on executions of an input generator implies reduction on generated test inputs.
There exist situations where such property does not hold. A quick example written in Python is shown as follows: 

\begin{lstlisting}[language=Python, label={lst:loop}]
    import random
    x = [1, 2]
    if random.choice([False, True]):
        x.pop() # Remove the last element from list x.
    print(x)
\end{lstlisting}

Given an execution on above generator which yields $I=\CodeIn{[1]}$: 
$$
E: \{\} \stackrel{\CodeIn{x=[1, 2]}} \longrightarrow
\{x = \CodeIn{[1, 2]}\} \stackrel{\CodeIn{if (True)}} \longrightarrow
\{x = \CodeIn{[1, 2]}\} \stackrel{\CodeIn{x.pop()}} \longrightarrow
\{x=\CodeIn{[1]}\} \stackrel{\CodeIn{print(x)}} \longrightarrow
\{output=\CodeIn{[1]}, ...\},
$$
\toolname{} will find another execution which yields $I^{\prime}=\CodeIn{[1, 2]}$ that has a larger size compared to $I$:
$$
E^{\prime}: \{\} \stackrel{\CodeIn{x = [1, 2]}} 
\longrightarrow
\{x = \CodeIn{[1, 2]}\} \stackrel{\CodeIn{if (False)}} \longrightarrow
\{x = \CodeIn{[1, 2]}\} 
\stackrel{\CodeIn{print(x)}} \longrightarrow
\{output=\CodeIn{[1, 2]}, ...\}.
$$
As demonstrated by this example, if semantics of some operations in the reducible parts of the trace correspond to ``removing/deleting some contents in the generator's output'', reduction on the execution by excluding these operations will result in an increase of generator's output. This outcome mismatches our goal of reduction on the test input that generator yields.

\textbf{Locality.}
$
E^{\prime} \approx E
\Longrightarrow
I^{\prime} \approx I.
$
If we do local search (as many greedy search strategies used) on reduction on execution, the locality is essential for the performance of the search. However, there exist situations where a continuous reduction on execution of generator does not imply a continuous changes of the given test input.
To illustrate this concept, consider an extreme example: if the generator outputs a hash value of a randomly generated string, preserving locality is nearly impossible. A concrete example is as follows (as one of its execution traces $E$ and derived $E^{\prime}$ are also shown):

\begin{lstlisting}[language=Python, label={lst:loop}]
    import random
    import hashlib
    s = ""
    for _ in range(random.choice(range(10))):
        s += "a"
    print(hashlib.md5(s.encode("utf-8")).hexdigest())
\end{lstlisting}
\begin{equation*}
    \begin{aligned}
&E: \{\} \stackrel{\CodeIn{s=""}} \longrightarrow
\{s=\CodeIn{""}\} \stackrel{\CodeIn{s+="a"}} \longrightarrow
\{s=\CodeIn{"a"}\} \stackrel{\CodeIn{s+="a"}} \longrightarrow
\{s=\CodeIn{"aa"}\} \stackrel{\CodeIn{print(...)}} \longrightarrow
\{output=\CodeIn{"4124bc0a..."}, ...\}
\\
&E^{\prime}: \{\} \stackrel{\CodeIn{s=""}} \longrightarrow
\{s=\CodeIn{""}\} \stackrel{\CodeIn{s+="a"}} \longrightarrow
\{s=\CodeIn{"a"}\} \stackrel{\CodeIn{print(...)}} \longrightarrow
\{output=\CodeIn{"0cc175b9..."}, ...\}      
    \end{aligned}
\end{equation*}

Luckily, in our preliminary study on generators collected in Zest~\cite{padhye2019semantic}, all of the generators preserve the monotonicity and locality. Furthermore, regarding the monotonicity, even in cases where the logic of generators corresponds to ``decreasing the output'' (which violates monotonicity), we can automatically identify or allow users to annotate them. We can then adjust our reduction strategy accordingly \pre{in}\rev{as a} future work.

\rev{
\rev{
\subsection{Limitations of Designs and Implementations}
}
To adapt our approach to a broader range of test generators, our designs and implementations still hold some limitations, as briefly demonstrated below.

\textbf{Limitation of \tone{}}.
Our design of reducible patterns in \tone{} is not for a reducer that holds with completeness. The trade-off is that we define a manageable search space that can effectively reduce real-world traces. One can extend this search space (such as bounded-exhaustive enumeration) to achieve better effectiveness in reduction, at the cost of significantly increased running time.
The evaluation in Section~\ref{sec:s5} shows that our design achieves good effectiveness and efficiency in practice. Also, we could design additional reducible patterns to cover more cases, our study suggests that the impact would be minor for most generators. Therefore, we leave this as a future work.

\textbf{Limitation of \ttwo{}}. 
When traces are unaligned (i.e., infeasible trace alignment described in Section~\ref{s4.4.1}) during execution, the strategies we designed in \ttwo{} (i.e., ``halt'', ``bypass'', and ``re-align'' strategies described in Section~\ref{s4.4.2}) will relieve this problem. These strategies tend to make the effort to reserve the semantics of remaining parts during execution. Nevertheless, these strategies are indeed insufficient to completely resolve the problem. For example, due to the complicated data dependencies in CSmith~\cite{DBLP:conf/pldi/YangCER11} (including the various customized data structures and program analysis such as point analysis and type inference), the strategies designed in our approach will be limited since the operations in the parts of infeasible trace alignment will affect the subsequent execution to a great extent.

\textbf{Limitation on Supported Generators}. The implementation of our approach currently only supports generators written in Java and Python, although our approach itself is general and not limited by the type of programming language. Due to the issue of implementation cost, we leave the support of our approach on other languages and frameworks as a future work.}
Another alternative way is to design a specific language for the generator to ensure \pre{these}\rev{above} good properties \rev{as well as address above limitations}. We have noticed some existing research works on the language design for generators such as Luck~\cite{luck}, Polyglot~\cite{chen2021one}, reflective generators~\cite{10.1145/3607842}, and Xsmith~\cite{Hatch23Generating}. How to design an appropriate language for writing input generators that adhere to \pre{these}\rev{above} desired properties \rev{and address above limitations} is also left as a future work. 

\section{Related Work}~\label{sec:relatedwork}
\subsection{Generator-based Testing}
Generator-based testing utilizes generator programs to produce test inputs that satisfy input specification.
As \pre{as}\rev{}a pioneering work, QuickCheck~\cite{DBLP:conf/icfp/ClaessenH00} first formulates tests as properties $\forall x: P(x) \Longrightarrow Q(x)$, which can be validated by generating multiple instances of $x$ that satisfy $P(x)$. The main challenge is to ensure the satisfaction of $P(x)$ for test inputs, necessitating efficient and effective generators. 

As a recent work, Zest~\cite{padhye2019semantic} uses a choice sequence model to parameterize the generated test. Based on junit-quickcheck~\cite{junit-quickcheck}, Zest builds many random generators for common objects such as graphs, bcel, maven configurations. Compared to fuzzing tools which generate object randomly, generators in Zest are able to improve the effectiveness of testing via ensuring the specification of test inputs by construction.
Our work is based on generator-based testing, aiming to reduce the generated test inputs that developer interests. The methodology of our work can be applied on various generators.

\rev{
Bonsai Fuzzing~\cite{vikram2021growing}, as another recent work, is a bounded generator-based testing technique. In order to obtain concise test cases that users need, it controls three properties of generators for bounding the generation process: maximum number of identifiers, maximum number of repetitions in any expansion of a Kleene-star, and maximum number of depth. Actually, the second property of their work is similar to the reducible patterns designed in our approach.
The main difference between Bonsai Fuzzing and our approach is that their application scenarios are different. Bonsai Fuzzing focuses on test generation for a corpus
of test inputs, as a contrast, our approach focuses on delta debugging, i.e., reducing a single given large test input. The underlying principles of how these two approaches work also differ significantly. Bonsai Fuzzing tends to incrementally enlarge the properties to gain test inputs users want. While our approach conducts search in a space defined by the reducible patterns we designed.
}

\subsection{Delta Debugging and Test Input Reduction}
Zeller and Hildebrandt 
initially proposed the delta debugging problem  ~\cite{ODD,DD}, as well as the first delta debugging algorithm, named \emph{ddmin}.
Following \emph{ddmin},
Misherghi and Su  
proposed Hierarchical Delta Debugging (HDD) ~\cite{HDD}, which utilizes the syntactical structure of the test input to improve the reduction process by applying ~\emph{ddmin} on the parse tree of the test input.
\pre{Sun et al. proposed Perses ~\cite{Perses} to further preserve the syntactical validity of the test input during the reduction process using a syntax-guided reduction algorithm.}
Wang et al. proposed ProbDD~\cite{ProbDD}, which introduces a probabilistic model to estimate the probability of each element to be kept in the produced result, and prioritizes the reduction of those with high probabilities. \citet{zhang2024deep} propose a refinement of ProbDD. \citet{wang2023probabilistic} further extend ProbDD to context-free languages.
\rev{
Sun et al. proposed Perses ~\cite{Perses} to preserve the syntactical validity of the test inputs during the reduction process. 
To reduce given inputs by their syntactical structures, Perses will normalize the given context-free grammar as its own normal form, which mainly includes three types of production rules: Kleene-Star Node (\CodeIn{A::=B*}), Kleene-Plus Node (\CodeIn{A::=B+}), and Optional Node (\CodeIn{A::=B?}). 
The design of our reducible patterns is inspired by Perses: the first two types of rules (Kleene-Star and Kleene-Plus) are similar to our Reducible Loop Pattern; and the third type (Optional) is similar to our Reducible Selection Pattern.
We extend their design to support our needs for manipulations on the trace.
}

In additional to delta debugging and its derived algorithms, 
many approaches have been proposed for test input reduction on specific domains. For program reduction,
CReduce~\cite{CReduce} is a transformation-based reducer that specifically designed to reduce C/C++ programs by applying well-crafted transformations on programs.
CHISEL~\cite{10.1145/3243734.3243838} also targets at reducing C programs, which considers the dependencies between elements in C programs, and accelerates the reduction by reinforcement learning.
Vulcan~\cite{10.1145/3586049} and Lampr~\cite{DBLP:journals/corr/abs-2312-13064} are two post-processing techniques which performs three aggressive language-agnostic transformations on the 1-minimal results produced by previous program reduction approaches.
Furthermore, T-Rec~\cite{10.1145/3690631}
leverages the lexical syntax of programming languages to reduce tokens of programs.
Besides program reduction, 
Binary Reduction~\cite{DBLP:conf/sigsoft/KalhaugeP19} and Generalized Binary Reduction~\cite{10.1145/3453483.3454091} are proposed to solve test reduction problem for dependency graphs in Java bytecode.
Zhou et al. proposed a debugging approach for microservice systems with parallel optimization~\cite{DBLP:journals/tsc/ZhouPXSJLD22}.
Niemetz and Biere proposed
ddSMT~\cite{niemetz2013ddsmt} for conducting delta debugging on SMT formulas.




Maciver and Donaldson proposed ``internal test-case reduction''~\cite{DBLP:conf/ecoop/MaciverD19}, a method that applies reduction directly on the sequence of random choices with shortlex order~\cite{Shortlex_order}.
However, simply reducing the sequence of random choices according to the shortlex order, without taking into account their impact on the generator's execution, will markedly expand the search space and encompass numerous test inputs that are of no interest to us\rev{, overall diminishing the effectiveness and efficiency of the reduction as shown in our evaluation (Section~\ref{s5.2.3})}.
Donaldson et al. ~\cite{10.1145/3453483.3454092} proposed an approach of test-case reduction and deduplication for SPIR-V programs in transformation-based testing.
This approach applies standard delta debugging to reduce a bug-inducing transformation sequence to a smaller subsequence, which shares similarities with the settings of $G^{\da}_{\ssb}$ in our experiments.
This approach is based on the assumption that transformations are designed to be as small and independent as possible. However,
for crafting test inputs with complex specification, generators may heavily have dependencies inside which introduce a challenge for the above approach. By contrast, our approach addresses this challenge with our proposed strategies. The evaluation results (comparing $G^{\da}_{\ssb}$ and $G^{\db}_{\ssc}$) confirm the superiority of our approach.



\subsection{Program Slicing}
Program slicing~\cite{weiser1979program} is a technique that computes the set of program statements that may affect the values at a specific point of interest, referred to as a slicing criterion. Program slicing is studied primarily in two forms: static~\cite{weiser1979program, harman2001overview} and dynamic~\cite{korel1988dynamic, agrawal1990dynamic}. 
In dynamic slicing, the program is executed on an input, and the resulting execution trace (for that input) is sliced, including only those parts that could have caused the fault to occurr during the particular execution of interest.

Program slicing can be used to aid the location of faults for software debugging.
The difference between program slicing and delta debugging lies in their focuses: program slicing focuses on source code analysis of the software under testing, while delta debugging conducts reduction on test inputs.
Inspired by the dynamic slicing technique on the trace of program execution~\cite{DBLP:conf/pldi/JhalaM05}, our approach conducts analysis and instrumentation on the test input generators, as well as utilizing the search strategies from existing delta debugging algorithm, to finally achieve reduction on test inputs.

\section{Conclusion}

In this paper, we proposed a novel approach named \toolname{} for test input reduction under generator-based testing. 
We conducted \toolname{} on three real-world application domains: graphs, DL models, and JavaScript programs.
We adopted \toolname{} on three generators from previous work. The evaluation shows that \toolname{} significantly outperforms the state-of-the-art reducer Perses and T-PDD in all domains: \toolname{}'s reduction results are 28.5\%, 34.6\%, 75.6\% in size of those from Perses and \toolname{} takes 17.5\%, 0.6\%, 65.4\% reduction time taken by Perses;
\toolname{}'s reduction results are 15.2\%, 8.8\%, 83.9\% in size of those from T-PDD and \toolname{} takes 13.9\%, 0.5\%, 22.0\% reduction time taken by T-PDD.
\rev{\toolname{} also substantially outperforms the state-of-the-art choice-sequence-based reduction approach named Hypothesis on the benchmarks from Hypothesis.
}
These results demonstrate effectiveness, efficiency and versatility of our approach.



\section{Acknowledgments}
This work was partially supported by National Natural Science Foundation of China under Grant No. 62161146003, and the Tencent Foundation/XPLORER PRIZE.
We would like to thank Xin Zhang, Chenxi Li for their help with improving the presentation of this work.


\clearpage








\newpage
\bibliographystyle{ACM-Reference-Format}
\bibliography{sample-base}


\begin{thebibliography}{60}


\ifx \showCODEN    \undefined \def \showCODEN     #1{\unskip}     \fi
\ifx \showDOI      \undefined \def \showDOI       #1{#1}\fi
\ifx \showISBNx    \undefined \def \showISBNx     #1{\unskip}     \fi
\ifx \showISBNxiii \undefined \def \showISBNxiii  #1{\unskip}     \fi
\ifx \showISSN     \undefined \def \showISSN      #1{\unskip}     \fi
\ifx \showLCCN     \undefined \def \showLCCN      #1{\unskip}     \fi
\ifx \shownote     \undefined \def \shownote      #1{#1}          \fi
\ifx \showarticletitle \undefined \def \showarticletitle #1{#1}   \fi
\ifx \showURL      \undefined \def \showURL       {\relax}        \fi
\providecommand\bibfield[2]{#2}
\providecommand\bibinfo[2]{#2}
\providecommand\natexlab[1]{#1}
\providecommand\showeprint[2][]{arXiv:#2}

\bibitem[\protect\citeauthoryear{Agrawal and Horgan}{Agrawal and
  Horgan}{1990}]%
        {agrawal1990dynamic}
\bibfield{author}{\bibinfo{person}{Hiralal Agrawal} {and}
  \bibinfo{person}{Joseph~R Horgan}.} \bibinfo{year}{1990}\natexlab{}.
\newblock \showarticletitle{Dynamic program slicing}.
\newblock \bibinfo{journal}{\emph{ACM SIGPlan Notices}} \bibinfo{volume}{25},
  \bibinfo{number}{6} (\bibinfo{year}{1990}), \bibinfo{pages}{246--256}.
\newblock


\bibitem[\protect\citeauthoryear{Chen, Moreau, Jiang, Zheng, Yan, Shen, Cowan,
  Wang, Hu, Ceze, Guestrin, and Krishnamurthy}{Chen et~al\mbox{.}}{2018}]%
        {DBLP:conf/osdi/ChenMJZYSCWHCGK18}
\bibfield{author}{\bibinfo{person}{Tianqi Chen}, \bibinfo{person}{Thierry
  Moreau}, \bibinfo{person}{Ziheng Jiang}, \bibinfo{person}{Lianmin Zheng},
  \bibinfo{person}{Eddie~Q. Yan}, \bibinfo{person}{Haichen Shen},
  \bibinfo{person}{Meghan Cowan}, \bibinfo{person}{Leyuan Wang},
  \bibinfo{person}{Yuwei Hu}, \bibinfo{person}{Luis Ceze},
  \bibinfo{person}{Carlos Guestrin}, {and} \bibinfo{person}{Arvind
  Krishnamurthy}.} \bibinfo{year}{2018}\natexlab{}.
\newblock \showarticletitle{{TVM:} An Automated End-to-End Optimizing Compiler
  for Deep Learning}. In \bibinfo{booktitle}{\emph{13th {USENIX} Symposium on
  Operating Systems Design and Implementation, {OSDI} 2018, Carlsbad, CA, USA,
  October 8-10, 2018}}, \bibfield{editor}{\bibinfo{person}{Andrea~C.
  Arpaci{-}Dusseau} {and} \bibinfo{person}{Geoff Voelker}} (Eds.).
  \bibinfo{publisher}{{USENIX} Association}, \bibinfo{pages}{578--594}.
\newblock
\urldef\tempurl%
\url{https://www.usenix.org/conference/osdi18/presentation/chen}
\showURL{%
\tempurl}


\bibitem[\protect\citeauthoryear{Chen, Zhong, Hu, Zhang, Yang, Wu, and
  Lee}{Chen et~al\mbox{.}}{2021}]%
        {chen2021one}
\bibfield{author}{\bibinfo{person}{Yongheng Chen}, \bibinfo{person}{Rui Zhong},
  \bibinfo{person}{Hong Hu}, \bibinfo{person}{Hangfan Zhang},
  \bibinfo{person}{Yupeng Yang}, \bibinfo{person}{Dinghao Wu}, {and}
  \bibinfo{person}{Wenke Lee}.} \bibinfo{year}{2021}\natexlab{}.
\newblock \showarticletitle{One engine to fuzz’em all: Generic language
  processor testing with semantic validation}. In
  \bibinfo{booktitle}{\emph{2021 IEEE Symposium on Security and Privacy (SP)}}.
  IEEE, \bibinfo{pages}{642--658}.
\newblock


\bibitem[\protect\citeauthoryear{Claessen and Hughes}{Claessen and
  Hughes}{2000}]%
        {DBLP:conf/icfp/ClaessenH00}
\bibfield{author}{\bibinfo{person}{Koen Claessen} {and} \bibinfo{person}{John
  Hughes}.} \bibinfo{year}{2000}\natexlab{}.
\newblock \showarticletitle{QuickCheck: a lightweight tool for random testing
  of Haskell programs}. In \bibinfo{booktitle}{\emph{Proceedings of the Fifth
  {ACM} {SIGPLAN} International Conference on Functional Programming {(ICFP}
  '00), Montreal, Canada, September 18-21, 2000}},
  \bibfield{editor}{\bibinfo{person}{Martin Odersky} {and}
  \bibinfo{person}{Philip Wadler}} (Eds.). \bibinfo{publisher}{{ACM}},
  \bibinfo{pages}{268--279}.
\newblock
\urldef\tempurl%
\url{https://doi.org/10.1145/351240.351266}
\showDOI{\tempurl}


\bibitem[\protect\citeauthoryear{difflib}{difflib}{2023}]%
        {difflib}
\bibfield{author}{\bibinfo{person}{difflib}.} \bibinfo{year}{2023}\natexlab{}.
\newblock \bibinfo{title}{difflib — Helpers for computing deltas.}
\newblock
\newblock
\urldef\tempurl%
\url{https://docs.python.org/3/library/difflib.html}
\showURL{%
\tempurl}
\newblock
\shownote{Accessed: 2024-02-01.}


\bibitem[\protect\citeauthoryear{Donaldson, Thomson, Teliman, Milizia, Maselco,
  and Karpi\'{n}ski}{Donaldson et~al\mbox{.}}{2021}]%
        {10.1145/3453483.3454092}
\bibfield{author}{\bibinfo{person}{Alastair~F. Donaldson},
  \bibinfo{person}{Paul Thomson}, \bibinfo{person}{Vasyl Teliman},
  \bibinfo{person}{Stefano Milizia}, \bibinfo{person}{Andr\'{e}~Perez Maselco},
  {and} \bibinfo{person}{Antoni Karpi\'{n}ski}.}
  \bibinfo{year}{2021}\natexlab{}.
\newblock \showarticletitle{Test-Case Reduction and Deduplication Almost for
  Free with Transformation-Based Compiler Testing}. In
  \bibinfo{booktitle}{\emph{Proceedings of the 42nd ACM SIGPLAN International
  Conference on Programming Language Design and Implementation}}
  \emph{(\bibinfo{series}{PLDI 2021})}. \bibinfo{publisher}{Association for
  Computing Machinery}, \bibinfo{address}{New York, NY, USA},
  \bibinfo{pages}{1017–1032}.
\newblock
\showISBNx{9781450383912}
\urldef\tempurl%
\url{https://doi.org/10.1145/3453483.3454092}
\showDOI{\tempurl}


\bibitem[\protect\citeauthoryear{Foundation}{Foundation}{2024}]%
        {PyAST}
\bibfield{author}{\bibinfo{person}{The Python~Software Foundation}.}
  \bibinfo{year}{2024}\natexlab{}.
\newblock \bibinfo{title}{ast — Abstract Syntax Trees}.
\newblock
\newblock
\urldef\tempurl%
\url{https://docs.python.org/3/library/ast.html}
\showURL{%
\tempurl}
\newblock
\shownote{Accessed: 2024-07-01.}


\bibitem[\protect\citeauthoryear{GCC}{GCC}{2022}]%
        {GCCDoc}
\bibfield{author}{\bibinfo{person}{GCC}.} \bibinfo{year}{2022}\natexlab{}.
\newblock \bibinfo{title}{How to Minimize Test Cases for Bugs.}
\newblock
\newblock
\urldef\tempurl%
\url{https://gcc.gnu.org/bugs/minimize.html}
\showURL{%
\tempurl}
\newblock
\shownote{Accessed: 2024-02-01.}


\bibitem[\protect\citeauthoryear{Goldstein, Frohlich, Wang, and
  Pierce}{Goldstein et~al\mbox{.}}{2023}]%
        {10.1145/3607842}
\bibfield{author}{\bibinfo{person}{Harrison Goldstein},
  \bibinfo{person}{Samantha Frohlich}, \bibinfo{person}{Meng Wang}, {and}
  \bibinfo{person}{Benjamin~C. Pierce}.} \bibinfo{year}{2023}\natexlab{}.
\newblock \showarticletitle{Reflecting on Random Generation}.
\newblock \bibinfo{journal}{\emph{Proc. ACM Program. Lang.}}
  \bibinfo{volume}{7}, \bibinfo{number}{ICFP}, Article \bibinfo{articleno}{200}
  (\bibinfo{date}{aug} \bibinfo{year}{2023}), \bibinfo{numpages}{34}~pages.
\newblock
\urldef\tempurl%
\url{https://doi.org/10.1145/3607842}
\showDOI{\tempurl}


\bibitem[\protect\citeauthoryear{Google}{Google}{2023}]%
        {Closure}
\bibfield{author}{\bibinfo{person}{Google}.} \bibinfo{year}{2023}\natexlab{}.
\newblock \bibinfo{title}{Google Closure Compiler.}
\newblock
\newblock
\urldef\tempurl%
\url{https://github.com/google/closure-compiler}
\showURL{%
\tempurl}
\newblock
\shownote{Accessed: 2024-02-01.}


\bibitem[\protect\citeauthoryear{Groce, Pinto, Azimi, and Mittal}{Groce
  et~al\mbox{.}}{2015}]%
        {TSTL}
\bibfield{author}{\bibinfo{person}{Alex Groce}, \bibinfo{person}{Jervis Pinto},
  \bibinfo{person}{Pooria Azimi}, {and} \bibinfo{person}{Pranjal Mittal}.}
  \bibinfo{year}{2015}\natexlab{}.
\newblock \showarticletitle{TSTL: a language and tool for testing (demo)}. In
  \bibinfo{booktitle}{\emph{Proceedings of the 2015 International Symposium on
  Software Testing and Analysis}} \emph{(\bibinfo{series}{ISSTA 2015})}.
  \bibinfo{publisher}{Association for Computing Machinery},
  \bibinfo{address}{New York, NY, USA}, \bibinfo{pages}{414–417}.
\newblock
\showISBNx{9781450336208}
\urldef\tempurl%
\url{https://doi.org/10.1145/2771783.2784769}
\showDOI{\tempurl}


\bibitem[\protect\citeauthoryear{Gu, Luo, Zhou, and Wang}{Gu
  et~al\mbox{.}}{2022}]%
        {DBLP:conf/icse/GuLZ022}
\bibfield{author}{\bibinfo{person}{Jiazhen Gu}, \bibinfo{person}{Xuchuan Luo},
  \bibinfo{person}{Yangfan Zhou}, {and} \bibinfo{person}{Xin Wang}.}
  \bibinfo{year}{2022}\natexlab{}.
\newblock \showarticletitle{Muffin: Testing Deep Learning Libraries via Neural
  Architecture Fuzzing}. In \bibinfo{booktitle}{\emph{44th {IEEE/ACM} 44th
  International Conference on Software Engineering, {ICSE} 2022, Pittsburgh,
  PA, USA, May 25-27, 2022}}. \bibinfo{publisher}{{ACM}},
  \bibinfo{pages}{1418--1430}.
\newblock
\urldef\tempurl%
\url{https://doi.org/10.1145/3510003.3510092}
\showDOI{\tempurl}


\bibitem[\protect\citeauthoryear{Harman and Hierons}{Harman and
  Hierons}{2001}]%
        {harman2001overview}
\bibfield{author}{\bibinfo{person}{Mark Harman} {and} \bibinfo{person}{Robert
  Hierons}.} \bibinfo{year}{2001}\natexlab{}.
\newblock \showarticletitle{An overview of program slicing}.
\newblock \bibinfo{journal}{\emph{software focus}} \bibinfo{volume}{2},
  \bibinfo{number}{3} (\bibinfo{year}{2001}), \bibinfo{pages}{85--92}.
\newblock


\bibitem[\protect\citeauthoryear{Hatch, Darragh, Porncharoenwase, Watson, and
  Eide}{Hatch et~al\mbox{.}}{2023}]%
        {Hatch23Generating}
\bibfield{author}{\bibinfo{person}{William~Gallard Hatch},
  \bibinfo{person}{Pierce Darragh}, \bibinfo{person}{Sorawee Porncharoenwase},
  \bibinfo{person}{Guy Watson}, {and} \bibinfo{person}{Eric Eide}.}
  \bibinfo{year}{2023}\natexlab{}.
\newblock \showarticletitle{Generating Conforming Programs With Xsmith}. In
  \bibinfo{booktitle}{\emph{GPCE}}.
\newblock
\urldef\tempurl%
\url{https://doi.org/10.1145/3624007.3624056}
\showDOI{\tempurl}


\bibitem[\protect\citeauthoryear{Heo, Lee, Pashakhanloo, and Naik}{Heo
  et~al\mbox{.}}{2018}]%
        {10.1145/3243734.3243838}
\bibfield{author}{\bibinfo{person}{Kihong Heo}, \bibinfo{person}{Woosuk Lee},
  \bibinfo{person}{Pardis Pashakhanloo}, {and} \bibinfo{person}{Mayur Naik}.}
  \bibinfo{year}{2018}\natexlab{}.
\newblock \showarticletitle{Effective Program Debloating via Reinforcement
  Learning}. In \bibinfo{booktitle}{\emph{Proceedings of the 2018 ACM SIGSAC
  Conference on Computer and Communications Security}}
  \emph{(\bibinfo{series}{CCS '18})}. \bibinfo{publisher}{Association for
  Computing Machinery}, \bibinfo{address}{New York, NY, USA},
  \bibinfo{pages}{380–394}.
\newblock
\showISBNx{9781450356930}
\urldef\tempurl%
\url{https://doi.org/10.1145/3243734.3243838}
\showDOI{\tempurl}


\bibitem[\protect\citeauthoryear{Holler, Herzig, and Zeller}{Holler
  et~al\mbox{.}}{2012}]%
        {holler2012fuzzing}
\bibfield{author}{\bibinfo{person}{Christian Holler}, \bibinfo{person}{Kim
  Herzig}, {and} \bibinfo{person}{Andreas Zeller}.}
  \bibinfo{year}{2012}\natexlab{}.
\newblock \showarticletitle{Fuzzing with code fragments}. In
  \bibinfo{booktitle}{\emph{21st USENIX Security Symposium (USENIX Security
  12)}}. \bibinfo{pages}{445--458}.
\newblock


\bibitem[\protect\citeauthoryear{Holser}{Holser}{2023}]%
        {junit-quickcheck}
\bibfield{author}{\bibinfo{person}{Paul Holser}.}
  \bibinfo{year}{2023}\natexlab{}.
\newblock \bibinfo{title}{junit-quickcheck: Property-based testing,
  JUnit-style.}
\newblock
\newblock
\urldef\tempurl%
\url{https://github.com/pholser/junit-quickcheck}
\showURL{%
\tempurl}
\newblock
\shownote{Accessed: 2024-02-01.}


\bibitem[\protect\citeauthoryear{Hua, Lin, Ren, Li, Zhang, Jiao, and Xie}{Hua
  et~al\mbox{.}}{2023}]%
        {DBLP:conf/issta/HuaLRLZJ023}
\bibfield{author}{\bibinfo{person}{Ziyue Hua}, \bibinfo{person}{Wei Lin},
  \bibinfo{person}{Luyao Ren}, \bibinfo{person}{Zongyang Li},
  \bibinfo{person}{Lu Zhang}, \bibinfo{person}{Wenpin Jiao}, {and}
  \bibinfo{person}{Tao Xie}.} \bibinfo{year}{2023}\natexlab{}.
\newblock \showarticletitle{GDsmith: Detecting Bugs in Cypher Graph Database
  Engines}. In \bibinfo{booktitle}{\emph{Proceedings of the 32nd {ACM}
  {SIGSOFT} International Symposium on Software Testing and Analysis, {ISSTA}
  2023, Seattle, WA, USA, July 17-21, 2023}},
  \bibfield{editor}{\bibinfo{person}{Ren{\'{e}} Just} {and}
  \bibinfo{person}{Gordon Fraser}} (Eds.). \bibinfo{publisher}{{ACM}},
  \bibinfo{pages}{163--174}.
\newblock
\urldef\tempurl%
\url{https://doi.org/10.1145/3597926.3598046}
\showDOI{\tempurl}


\bibitem[\protect\citeauthoryear{IBM}{IBM}{2017}]%
        {WALA}
\bibfield{author}{\bibinfo{person}{IBM}.} \bibinfo{year}{2017}\natexlab{}.
\newblock \bibinfo{title}{The T.J. Watson Libraries for Analysis.}
\newblock
\newblock
\urldef\tempurl%
\url{http://wala.sourceforge.net/}
\showURL{%
\tempurl}
\newblock
\shownote{Accessed: 2024-02-01.}


\bibitem[\protect\citeauthoryear{Jhala and Majumdar}{Jhala and
  Majumdar}{2005}]%
        {DBLP:conf/pldi/JhalaM05}
\bibfield{author}{\bibinfo{person}{Ranjit Jhala} {and} \bibinfo{person}{Rupak
  Majumdar}.} \bibinfo{year}{2005}\natexlab{}.
\newblock \showarticletitle{Path slicing}. In
  \bibinfo{booktitle}{\emph{Proceedings of the {ACM} {SIGPLAN} 2005 Conference
  on Programming Language Design and Implementation, Chicago, IL, USA, June
  12-15, 2005}}, \bibfield{editor}{\bibinfo{person}{Vivek Sarkar} {and}
  \bibinfo{person}{Mary~W. Hall}} (Eds.). \bibinfo{publisher}{{ACM}},
  \bibinfo{pages}{38--47}.
\newblock
\urldef\tempurl%
\url{https://doi.org/10.1145/1065010.1065016}
\showDOI{\tempurl}


\bibitem[\protect\citeauthoryear{Kalhauge and Palsberg}{Kalhauge and
  Palsberg}{2019}]%
        {DBLP:conf/sigsoft/KalhaugeP19}
\bibfield{author}{\bibinfo{person}{Christian~Gram Kalhauge} {and}
  \bibinfo{person}{Jens Palsberg}.} \bibinfo{year}{2019}\natexlab{}.
\newblock \showarticletitle{Binary reduction of dependency graphs}. In
  \bibinfo{booktitle}{\emph{Proceedings of the {ACM} Joint Meeting on European
  Software Engineering Conference and Symposium on the Foundations of Software
  Engineering, {ESEC/SIGSOFT} {FSE} 2019, Tallinn, Estonia, August 26-30,
  2019}}, \bibfield{editor}{\bibinfo{person}{Marlon Dumas},
  \bibinfo{person}{Dietmar Pfahl}, \bibinfo{person}{Sven Apel}, {and}
  \bibinfo{person}{Alessandra Russo}} (Eds.). \bibinfo{publisher}{{ACM}},
  \bibinfo{pages}{556--566}.
\newblock
\urldef\tempurl%
\url{https://doi.org/10.1145/3338906.3338956}
\showDOI{\tempurl}


\bibitem[\protect\citeauthoryear{Kalhauge and Palsberg}{Kalhauge and
  Palsberg}{2021}]%
        {10.1145/3453483.3454091}
\bibfield{author}{\bibinfo{person}{Christian~Gram Kalhauge} {and}
  \bibinfo{person}{Jens Palsberg}.} \bibinfo{year}{2021}\natexlab{}.
\newblock \showarticletitle{Logical bytecode reduction}. In
  \bibinfo{booktitle}{\emph{Proceedings of the 42nd ACM SIGPLAN International
  Conference on Programming Language Design and Implementation}}
  \emph{(\bibinfo{series}{PLDI 2021})}. \bibinfo{publisher}{Association for
  Computing Machinery}, \bibinfo{address}{New York, NY, USA},
  \bibinfo{pages}{1003–1016}.
\newblock
\showISBNx{9781450383912}
\urldef\tempurl%
\url{https://doi.org/10.1145/3453483.3454091}
\showDOI{\tempurl}


\bibitem[\protect\citeauthoryear{Karp}{Karp}{1972}]%
        {DBLP:conf/coco/Karp72}
\bibfield{author}{\bibinfo{person}{Richard~M. Karp}.}
  \bibinfo{year}{1972}\natexlab{}.
\newblock \showarticletitle{Reducibility Among Combinatorial Problems}. In
  \bibinfo{booktitle}{\emph{Proceedings of a symposium on the Complexity of
  Computer Computations, held March 20-22, 1972, at the {IBM} Thomas J. Watson
  Research Center, Yorktown Heights, New York, {USA}}}
  \emph{(\bibinfo{series}{The {IBM} Research Symposia Series})},
  \bibfield{editor}{\bibinfo{person}{Raymond~E. Miller} {and}
  \bibinfo{person}{James~W. Thatcher}} (Eds.). \bibinfo{publisher}{Plenum
  Press, New York}, \bibinfo{pages}{85--103}.
\newblock
\urldef\tempurl%
\url{https://doi.org/10.1007/978-1-4684-2001-2\_9}
\showDOI{\tempurl}


\bibitem[\protect\citeauthoryear{Korel and Laski}{Korel and Laski}{1988}]%
        {korel1988dynamic}
\bibfield{author}{\bibinfo{person}{Bogdan Korel} {and} \bibinfo{person}{Janusz
  Laski}.} \bibinfo{year}{1988}\natexlab{}.
\newblock \showarticletitle{Dynamic program slicing}.
\newblock \bibinfo{journal}{\emph{Information processing letters}}
  \bibinfo{volume}{29}, \bibinfo{number}{3} (\bibinfo{year}{1988}),
  \bibinfo{pages}{155--163}.
\newblock


\bibitem[\protect\citeauthoryear{Kukucka, Pina, Ammann, and Bell}{Kukucka
  et~al\mbox{.}}{2022}]%
        {kukucka2022confetti}
\bibfield{author}{\bibinfo{person}{James Kukucka}, \bibinfo{person}{Lu{\'\i}s
  Pina}, \bibinfo{person}{Paul Ammann}, {and} \bibinfo{person}{Jonathan Bell}.}
  \bibinfo{year}{2022}\natexlab{}.
\newblock \showarticletitle{Confetti: Amplifying concolic guidance for
  fuzzers}. In \bibinfo{booktitle}{\emph{Proceedings of the 44th International
  Conference on Software Engineering}}. \bibinfo{pages}{438--450}.
\newblock


\bibitem[\protect\citeauthoryear{Lampropoulos, Gallois-Wong, Hri{\c{t}}cu,
  Hughes, Pierce, and Xia}{Lampropoulos et~al\mbox{.}}{2017}]%
        {luck}
\bibfield{author}{\bibinfo{person}{Leonidas Lampropoulos},
  \bibinfo{person}{Diane Gallois-Wong}, \bibinfo{person}{C{\u{a}}t{\u{a}}lin
  Hri{\c{t}}cu}, \bibinfo{person}{John Hughes}, \bibinfo{person}{Benjamin~C
  Pierce}, {and} \bibinfo{person}{Li-yao Xia}.}
  \bibinfo{year}{2017}\natexlab{}.
\newblock \showarticletitle{Beginner's luck: a language for property-based
  generators}. In \bibinfo{booktitle}{\emph{Proceedings of the 44th ACM SIGPLAN
  Symposium on Principles of Programming Languages}}.
  \bibinfo{pages}{114--129}.
\newblock


\bibitem[\protect\citeauthoryear{Livinskii, Babokin, and Regehr}{Livinskii
  et~al\mbox{.}}{2020}]%
        {10.1145/3428264}
\bibfield{author}{\bibinfo{person}{Vsevolod Livinskii}, \bibinfo{person}{Dmitry
  Babokin}, {and} \bibinfo{person}{John Regehr}.}
  \bibinfo{year}{2020}\natexlab{}.
\newblock \showarticletitle{Random Testing for C and C++ Compilers with
  YARPGen}.
\newblock \bibinfo{journal}{\emph{Proc. ACM Program. Lang.}}
  \bibinfo{volume}{4}, \bibinfo{number}{OOPSLA}, Article
  \bibinfo{articleno}{196} (\bibinfo{date}{nov} \bibinfo{year}{2020}),
  \bibinfo{numpages}{25}~pages.
\newblock
\urldef\tempurl%
\url{https://doi.org/10.1145/3428264}
\showDOI{\tempurl}


\bibitem[\protect\citeauthoryear{Maciver and Donaldson}{Maciver and
  Donaldson}{2020}]%
        {DBLP:conf/ecoop/MaciverD19}
\bibfield{author}{\bibinfo{person}{David Maciver} {and}
  \bibinfo{person}{Alastair~F. Donaldson}.} \bibinfo{year}{2020}\natexlab{}.
\newblock \showarticletitle{Test-Case Reduction via Test-Case Generation:
  Insights from the Hypothesis Reducer (Tool Insights Paper)}. In
  \bibinfo{booktitle}{\emph{34th European Conference on Object-Oriented
  Programming, {ECOOP} 2020, November 15-17, 2020, Berlin, Germany (Virtual
  Conference)}} \emph{(\bibinfo{series}{LIPIcs})},
  \bibfield{editor}{\bibinfo{person}{Robert Hirschfeld} {and}
  \bibinfo{person}{Tobias Pape}} (Eds.), Vol.~\bibinfo{volume}{166}.
  \bibinfo{publisher}{Schloss Dagstuhl - Leibniz-Zentrum f{\"{u}}r Informatik},
  \bibinfo{pages}{13:1--13:27}.
\newblock
\urldef\tempurl%
\url{https://doi.org/10.4230/LIPIcs.ECOOP.2020.13}
\showDOI{\tempurl}


\bibitem[\protect\citeauthoryear{Martin~Torp}{Martin~Torp}{2023}]%
        {JSDelta}
\bibfield{author}{\bibinfo{person}{Esben Sparre~Andreasen Martin~Torp}.}
  \bibinfo{year}{2023}\natexlab{}.
\newblock \bibinfo{title}{JS Delta.}
\newblock
\newblock
\urldef\tempurl%
\url{https://github.com/wala/jsdelta}
\showURL{%
\tempurl}
\newblock
\shownote{Accessed: 2024-02-01.}


\bibitem[\protect\citeauthoryear{Michail, Kinable, Naveh, and Sichi}{Michail
  et~al\mbox{.}}{2020}]%
        {10.1145/3381449}
\bibfield{author}{\bibinfo{person}{Dimitrios Michail}, \bibinfo{person}{Joris
  Kinable}, \bibinfo{person}{Barak Naveh}, {and} \bibinfo{person}{John~V.
  Sichi}.} \bibinfo{year}{2020}\natexlab{}.
\newblock \showarticletitle{JGraphT—A Java Library for Graph Data Structures
  and Algorithms}.
\newblock \bibinfo{journal}{\emph{ACM Trans. Math. Softw.}}
  \bibinfo{volume}{46}, \bibinfo{number}{2}, Article \bibinfo{articleno}{16}
  (\bibinfo{date}{may} \bibinfo{year}{2020}), \bibinfo{numpages}{29}~pages.
\newblock
\showISSN{0098-3500}
\urldef\tempurl%
\url{https://doi.org/10.1145/3381449}
\showDOI{\tempurl}


\bibitem[\protect\citeauthoryear{Mircrosoft}{Mircrosoft}{2017}]%
        {ONNX}
\bibfield{author}{\bibinfo{person}{Mircrosoft}.}
  \bibinfo{year}{2017}\natexlab{}.
\newblock \bibinfo{title}{Open Neural Network Exchange.}
\newblock
\newblock
\urldef\tempurl%
\url{https://onnx.ai/}
\showURL{%
\tempurl}
\newblock
\shownote{Accessed: 2024-02-01.}


\bibitem[\protect\citeauthoryear{Misherghi and Su}{Misherghi and Su}{2006}]%
        {HDD}
\bibfield{author}{\bibinfo{person}{Ghassan Misherghi} {and}
  \bibinfo{person}{Zhendong Su}.} \bibinfo{year}{2006}\natexlab{}.
\newblock \showarticletitle{{HDD:} hierarchical Delta Debugging}. In
  \bibinfo{booktitle}{\emph{28th International Conference on Software
  Engineering {(ICSE} 2006), Shanghai, China, May 20-28, 2006}},
  \bibfield{editor}{\bibinfo{person}{Leon~J. Osterweil},
  \bibinfo{person}{H.~Dieter Rombach}, {and} \bibinfo{person}{Mary~Lou Soffa}}
  (Eds.). \bibinfo{publisher}{{ACM}}, \bibinfo{pages}{142--151}.
\newblock
\urldef\tempurl%
\url{https://doi.org/10.1145/1134285.1134307}
\showDOI{\tempurl}


\bibitem[\protect\citeauthoryear{Nguyen and Grunske}{Nguyen and
  Grunske}{2022}]%
        {nguyen2022bedivfuzz}
\bibfield{author}{\bibinfo{person}{Hoang~Lam Nguyen} {and}
  \bibinfo{person}{Lars Grunske}.} \bibinfo{year}{2022}\natexlab{}.
\newblock \showarticletitle{BeDivFuzz: integrating behavioral diversity into
  generator-based fuzzing}. In \bibinfo{booktitle}{\emph{Proceedings of the
  44th International Conference on Software Engineering}}.
  \bibinfo{pages}{249--261}.
\newblock


\bibitem[\protect\citeauthoryear{Niemetz and Biere}{Niemetz and Biere}{2013}]%
        {niemetz2013ddsmt}
\bibfield{author}{\bibinfo{person}{Aina Niemetz} {and} \bibinfo{person}{Armin
  Biere}.} \bibinfo{year}{2013}\natexlab{}.
\newblock \showarticletitle{ddSMT: a delta debugger for the SMT-LIB v2 format}.
  In \bibinfo{booktitle}{\emph{Proceedings of the 11th International Workshop
  on Satisfiability Modulo Theories, SMT}}. \bibinfo{pages}{8--9}.
\newblock


\bibitem[\protect\citeauthoryear{Padhye, Lemieux, Sen, Papadakis, and
  Le~Traon}{Padhye et~al\mbox{.}}{2019}]%
        {padhye2019semantic}
\bibfield{author}{\bibinfo{person}{Rohan Padhye}, \bibinfo{person}{Caroline
  Lemieux}, \bibinfo{person}{Koushik Sen}, \bibinfo{person}{Mike Papadakis},
  {and} \bibinfo{person}{Yves Le~Traon}.} \bibinfo{year}{2019}\natexlab{}.
\newblock \showarticletitle{Semantic fuzzing with zest}. In
  \bibinfo{booktitle}{\emph{Proceedings of the 28th ACM SIGSOFT International
  Symposium on Software Testing and Analysis}}. \bibinfo{pages}{329--340}.
\newblock


\bibitem[\protect\citeauthoryear{Parr}{Parr}{2013}]%
        {ANTLR}
\bibfield{author}{\bibinfo{person}{Terence Parr}.}
  \bibinfo{year}{2013}\natexlab{}.
\newblock \showarticletitle{The definitive ANTLR 4 reference}.
\newblock \bibinfo{journal}{\emph{The Definitive ANTLR 4 Reference}}
  (\bibinfo{year}{2013}), \bibinfo{pages}{1--326}.
\newblock


\bibitem[\protect\citeauthoryear{Parr}{Parr}{2023}]%
        {ANTLR_repo}
\bibfield{author}{\bibinfo{person}{Terence Parr}.}
  \bibinfo{year}{2023}\natexlab{}.
\newblock \bibinfo{title}{ANTLR v4.}
\newblock
\newblock
\urldef\tempurl%
\url{https://github.com/antlr/antlr4}
\showURL{%
\tempurl}
\newblock
\shownote{Accessed: 2024-02-01.}


\bibitem[\protect\citeauthoryear{Pike}{Pike}{2014}]%
        {pike2014smartcheck}
\bibfield{author}{\bibinfo{person}{Lee Pike}.} \bibinfo{year}{2014}\natexlab{}.
\newblock \showarticletitle{SmartCheck: automatic and efficient counterexample
  reduction and generalization}. In \bibinfo{booktitle}{\emph{Proceedings of
  the 2014 ACM SIGPLAN symposium on Haskell}}. \bibinfo{pages}{53--64}.
\newblock


\bibitem[\protect\citeauthoryear{Reddy, Lemieux, Padhye, and Sen}{Reddy
  et~al\mbox{.}}{2020}]%
        {reddy2020quickly}
\bibfield{author}{\bibinfo{person}{Sameer Reddy}, \bibinfo{person}{Caroline
  Lemieux}, \bibinfo{person}{Rohan Padhye}, {and} \bibinfo{person}{Koushik
  Sen}.} \bibinfo{year}{2020}\natexlab{}.
\newblock \showarticletitle{Quickly generating diverse valid test inputs with
  reinforcement learning}. In \bibinfo{booktitle}{\emph{Proceedings of the
  ACM/IEEE 42nd International Conference on Software Engineering}}.
  \bibinfo{pages}{1410--1421}.
\newblock


\bibitem[\protect\citeauthoryear{Regehr, Chen, Cuoq, Eide, Ellison, and
  Yang}{Regehr et~al\mbox{.}}{2012}]%
        {CReduce}
\bibfield{author}{\bibinfo{person}{John Regehr}, \bibinfo{person}{Yang Chen},
  \bibinfo{person}{Pascal Cuoq}, \bibinfo{person}{Eric Eide},
  \bibinfo{person}{Chucky Ellison}, {and} \bibinfo{person}{Xuejun Yang}.}
  \bibinfo{year}{2012}\natexlab{}.
\newblock \showarticletitle{Test-case reduction for {C} compiler bugs}. In
  \bibinfo{booktitle}{\emph{{ACM} {SIGPLAN} Conference on Programming Language
  Design and Implementation, {PLDI} '12, Beijing, China - June 11 - 16, 2012}},
  \bibfield{editor}{\bibinfo{person}{Jan Vitek}, \bibinfo{person}{Haibo Lin},
  {and} \bibinfo{person}{Frank Tip}} (Eds.). \bibinfo{publisher}{{ACM}},
  \bibinfo{pages}{335--346}.
\newblock
\urldef\tempurl%
\url{https://doi.org/10.1145/2254064.2254104}
\showDOI{\tempurl}


\bibitem[\protect\citeauthoryear{Ren}{Ren}{2023}]%
        {projectwebsite}
\bibfield{author}{\bibinfo{person}{Luyao Ren}.}
  \bibinfo{year}{2023}\natexlab{}.
\newblock \bibinfo{title}{Project Website}.
\newblock
\newblock
\urldef\tempurl%
\url{https://github.com/GReduce/GReduce}
\showURL{%
\tempurl}
\newblock
\shownote{Accessed: 2024-02-01.}


\bibitem[\protect\citeauthoryear{Ren, Wang, Xiong, Zhang, Jiang, and Xie}{Ren
  et~al\mbox{.}}{2023}]%
        {DBLP:journals/corr/abs-2302-00842}
\bibfield{author}{\bibinfo{person}{Luyao Ren}, \bibinfo{person}{Ziheng Wang},
  \bibinfo{person}{Yingfei Xiong}, \bibinfo{person}{Li Zhang},
  \bibinfo{person}{Guoyue Jiang}, {and} \bibinfo{person}{Tao Xie}.}
  \bibinfo{year}{2023}\natexlab{}.
\newblock \showarticletitle{Effective Random Test Generation for Deep Learning
  Compilers}.
\newblock \bibinfo{journal}{\emph{CoRR}}  \bibinfo{volume}{abs/2302.00842}
  (\bibinfo{year}{2023}).
\newblock
\urldef\tempurl%
\url{https://doi.org/10.48550/arXiv.2302.00842}
\showDOI{\tempurl}
\showeprint[arXiv]{2302.00842}


\bibitem[\protect\citeauthoryear{Rendel and Ostermann}{Rendel and
  Ostermann}{2010}]%
        {rendel2010invertible}
\bibfield{author}{\bibinfo{person}{Tillmann Rendel} {and}
  \bibinfo{person}{Klaus Ostermann}.} \bibinfo{year}{2010}\natexlab{}.
\newblock \showarticletitle{Invertible syntax descriptions: unifying parsing
  and pretty printing}.
\newblock \bibinfo{journal}{\emph{ACM Sigplan Notices}} \bibinfo{volume}{45},
  \bibinfo{number}{11} (\bibinfo{year}{2010}), \bibinfo{pages}{1--12}.
\newblock


\bibitem[\protect\citeauthoryear{Sun, Li, Zhang, Gu, and Su}{Sun
  et~al\mbox{.}}{2018}]%
        {Perses}
\bibfield{author}{\bibinfo{person}{Chengnian Sun}, \bibinfo{person}{Yuanbo Li},
  \bibinfo{person}{Qirun Zhang}, \bibinfo{person}{Tianxiao Gu}, {and}
  \bibinfo{person}{Zhendong Su}.} \bibinfo{year}{2018}\natexlab{}.
\newblock \showarticletitle{Perses: syntax-guided program reduction}. In
  \bibinfo{booktitle}{\emph{Proceedings of the 40th International Conference on
  Software Engineering, {ICSE} 2018, Gothenburg, Sweden, May 27 - June 03,
  2018}}, \bibfield{editor}{\bibinfo{person}{Michel Chaudron},
  \bibinfo{person}{Ivica Crnkovic}, \bibinfo{person}{Marsha Chechik}, {and}
  \bibinfo{person}{Mark Harman}} (Eds.). \bibinfo{publisher}{{ACM}},
  \bibinfo{pages}{361--371}.
\newblock
\urldef\tempurl%
\url{https://doi.org/10.1145/3180155.3180236}
\showDOI{\tempurl}


\bibitem[\protect\citeauthoryear{Vikram, Padhye, and Sen}{Vikram
  et~al\mbox{.}}{2021}]%
        {vikram2021growing}
\bibfield{author}{\bibinfo{person}{Vasudev Vikram}, \bibinfo{person}{Rohan
  Padhye}, {and} \bibinfo{person}{Koushik Sen}.}
  \bibinfo{year}{2021}\natexlab{}.
\newblock \showarticletitle{Growing a test corpus with bonsai fuzzing}. In
  \bibinfo{booktitle}{\emph{2021 IEEE/ACM 43rd International Conference on
  Software Engineering (ICSE)}}. IEEE, \bibinfo{pages}{723--735}.
\newblock


\bibitem[\protect\citeauthoryear{Wang, Shen, Chen, Xiong, and Zhang}{Wang
  et~al\mbox{.}}{2021}]%
        {ProbDD}
\bibfield{author}{\bibinfo{person}{Guancheng Wang}, \bibinfo{person}{Ruobing
  Shen}, \bibinfo{person}{Junjie Chen}, \bibinfo{person}{Yingfei Xiong}, {and}
  \bibinfo{person}{Lu Zhang}.} \bibinfo{year}{2021}\natexlab{}.
\newblock \showarticletitle{Probabilistic Delta debugging}. In
  \bibinfo{booktitle}{\emph{{ESEC/FSE} '21: 29th {ACM} Joint European Software
  Engineering Conference and Symposium on the Foundations of Software
  Engineering, Athens, Greece, August 23-28, 2021}},
  \bibfield{editor}{\bibinfo{person}{Diomidis Spinellis},
  \bibinfo{person}{Georgios Gousios}, \bibinfo{person}{Marsha Chechik}, {and}
  \bibinfo{person}{Massimiliano~Di Penta}} (Eds.). \bibinfo{publisher}{{ACM}},
  \bibinfo{pages}{881--892}.
\newblock
\urldef\tempurl%
\url{https://doi.org/10.1145/3468264.3468625}
\showDOI{\tempurl}


\bibitem[\protect\citeauthoryear{Wang, Wu, Zhu, Xiong, Zhang, and Zhang}{Wang
  et~al\mbox{.}}{2023}]%
        {wang2023probabilistic}
\bibfield{author}{\bibinfo{person}{Guancheng Wang}, \bibinfo{person}{Yiqian
  Wu}, \bibinfo{person}{Qihao Zhu}, \bibinfo{person}{Yingfei Xiong},
  \bibinfo{person}{Xin Zhang}, {and} \bibinfo{person}{Lu Zhang}.}
  \bibinfo{year}{2023}\natexlab{}.
\newblock \showarticletitle{A Probabilistic Delta Debugging Approach for
  Abstract Syntax Trees}. In \bibinfo{booktitle}{\emph{2023 IEEE 34th
  International Symposium on Software Reliability Engineering (ISSRE)}}. IEEE,
  \bibinfo{pages}{763--773}.
\newblock


\bibitem[\protect\citeauthoryear{Wang, Chen, Wei, and Liu}{Wang
  et~al\mbox{.}}{2017}]%
        {DBLP:conf/sp/WangCWL17}
\bibfield{author}{\bibinfo{person}{Junjie Wang}, \bibinfo{person}{Bihuan Chen},
  \bibinfo{person}{Lei Wei}, {and} \bibinfo{person}{Yang Liu}.}
  \bibinfo{year}{2017}\natexlab{}.
\newblock \showarticletitle{Skyfire: Data-Driven Seed Generation for Fuzzing}.
  In \bibinfo{booktitle}{\emph{2017 {IEEE} Symposium on Security and Privacy,
  {SP} 2017, San Jose, CA, USA, May 22-26, 2017}}. \bibinfo{publisher}{{IEEE}
  Computer Society}, \bibinfo{pages}{579--594}.
\newblock
\urldef\tempurl%
\url{https://doi.org/10.1109/SP.2017.23}
\showDOI{\tempurl}


\bibitem[\protect\citeauthoryear{Weiser}{Weiser}{1979}]%
        {weiser1979program}
\bibfield{author}{\bibinfo{person}{Mark~David Weiser}.}
  \bibinfo{year}{1979}\natexlab{}.
\newblock \bibinfo{booktitle}{\emph{Program slices: formal, psychological, and
  practical investigations of an automatic program abstraction method}}.
\newblock \bibinfo{publisher}{University of Michigan}.
\newblock


\bibitem[\protect\citeauthoryear{Wikipedia}{Wikipedia}{2023a}]%
        {ins}
\bibfield{author}{\bibinfo{person}{Wikipedia}.}
  \bibinfo{year}{2023}\natexlab{a}.
\newblock \bibinfo{title}{Instrumentation (computer programming)}.
\newblock
\newblock
\urldef\tempurl%
\url{https://en.wikipedia.org/wiki/Instrumentation_(computer_programming)}
\showURL{%
\tempurl}
\newblock
\shownote{Accessed: 2024-02-01.}


\bibitem[\protect\citeauthoryear{Wikipedia}{Wikipedia}{2023b}]%
        {Shortlex_order}
\bibfield{author}{\bibinfo{person}{Wikipedia}.}
  \bibinfo{year}{2023}\natexlab{b}.
\newblock \bibinfo{title}{Shortlex order}.
\newblock
\newblock
\urldef\tempurl%
\url{https://en.wikipedia.org/wiki/Shortlex_order}
\showURL{%
\tempurl}
\newblock
\shownote{Accessed: 2024-02-01.}


\bibitem[\protect\citeauthoryear{Xin, Sumner, and Zhang}{Xin
  et~al\mbox{.}}{2008}]%
        {DBLP:conf/pldi/XinSZ08}
\bibfield{author}{\bibinfo{person}{Bin Xin}, \bibinfo{person}{William~N.
  Sumner}, {and} \bibinfo{person}{Xiangyu Zhang}.}
  \bibinfo{year}{2008}\natexlab{}.
\newblock \showarticletitle{Efficient program execution indexing}. In
  \bibinfo{booktitle}{\emph{Proceedings of the {ACM} {SIGPLAN} 2008 Conference
  on Programming Language Design and Implementation, Tucson, AZ, USA, June
  7-13, 2008}}, \bibfield{editor}{\bibinfo{person}{Rajiv Gupta} {and}
  \bibinfo{person}{Saman~P. Amarasinghe}} (Eds.). \bibinfo{publisher}{{ACM}},
  \bibinfo{pages}{238--248}.
\newblock
\urldef\tempurl%
\url{https://doi.org/10.1145/1375581.1375611}
\showDOI{\tempurl}


\bibitem[\protect\citeauthoryear{Xu, Tian, Zhang, Zhang, Liu, Jiang, and
  Sun}{Xu et~al\mbox{.}}{2024}]%
        {10.1145/3690631}
\bibfield{author}{\bibinfo{person}{Zhenyang Xu}, \bibinfo{person}{Yongqiang
  Tian}, \bibinfo{person}{Mengxiao Zhang}, \bibinfo{person}{Jiarui Zhang},
  \bibinfo{person}{Puzhuo Liu}, \bibinfo{person}{Yu Jiang}, {and}
  \bibinfo{person}{Chengnian Sun}.} \bibinfo{year}{2024}\natexlab{}.
\newblock \showarticletitle{T-Rec: Fine-Grained Language-Agnostic Program
  Reduction Guided by Lexical Syntax}.
\newblock \bibinfo{journal}{\emph{ACM Trans. Softw. Eng. Methodol.}}
  (\bibinfo{date}{Aug.} \bibinfo{year}{2024}).
\newblock
\showISSN{1049-331X}
\urldef\tempurl%
\url{https://doi.org/10.1145/3690631}
\showDOI{\tempurl}
\newblock
\shownote{Just Accepted.}


\bibitem[\protect\citeauthoryear{Xu, Tian, Zhang, Zhao, Jiang, and Sun}{Xu
  et~al\mbox{.}}{2023}]%
        {10.1145/3586049}
\bibfield{author}{\bibinfo{person}{Zhenyang Xu}, \bibinfo{person}{Yongqiang
  Tian}, \bibinfo{person}{Mengxiao Zhang}, \bibinfo{person}{Gaosen Zhao},
  \bibinfo{person}{Yu Jiang}, {and} \bibinfo{person}{Chengnian Sun}.}
  \bibinfo{year}{2023}\natexlab{}.
\newblock \showarticletitle{Pushing the Limit of 1-Minimality of
  Language-Agnostic Program Reduction}.
\newblock \bibinfo{journal}{\emph{Proc. ACM Program. Lang.}}
  \bibinfo{volume}{7}, \bibinfo{number}{OOPSLA1}, Article
  \bibinfo{articleno}{97} (\bibinfo{date}{apr} \bibinfo{year}{2023}),
  \bibinfo{numpages}{29}~pages.
\newblock
\urldef\tempurl%
\url{https://doi.org/10.1145/3586049}
\showDOI{\tempurl}


\bibitem[\protect\citeauthoryear{Yang, Chen, Eide, and Regehr}{Yang
  et~al\mbox{.}}{2011}]%
        {DBLP:conf/pldi/YangCER11}
\bibfield{author}{\bibinfo{person}{Xuejun Yang}, \bibinfo{person}{Yang Chen},
  \bibinfo{person}{Eric Eide}, {and} \bibinfo{person}{John Regehr}.}
  \bibinfo{year}{2011}\natexlab{}.
\newblock \showarticletitle{Finding and understanding bugs in {C} compilers}.
  In \bibinfo{booktitle}{\emph{Proceedings of the 32nd {ACM} {SIGPLAN}
  Conference on Programming Language Design and Implementation, {PLDI} 2011,
  San Jose, CA, USA, June 4-8, 2011}},
  \bibfield{editor}{\bibinfo{person}{Mary~W. Hall} {and}
  \bibinfo{person}{David~A. Padua}} (Eds.). \bibinfo{publisher}{{ACM}},
  \bibinfo{pages}{283--294}.
\newblock
\urldef\tempurl%
\url{https://doi.org/10.1145/1993498.1993532}
\showDOI{\tempurl}


\bibitem[\protect\citeauthoryear{Zeller}{Zeller}{1999}]%
        {ODD}
\bibfield{author}{\bibinfo{person}{Andreas Zeller}.}
  \bibinfo{year}{1999}\natexlab{}.
\newblock \showarticletitle{Yesterday, My Program Worked. Today, It Does Not.
  Why?}. In \bibinfo{booktitle}{\emph{Software Engineering - ESEC/FSE'99, 7th
  European Software Engineering Conference, Held Jointly with the 7th {ACM}
  {SIGSOFT} Symposium on the Foundations of Software Engineering, Toulouse,
  France, September 1999, Proceedings}} \emph{(\bibinfo{series}{Lecture Notes
  in Computer Science})}, \bibfield{editor}{\bibinfo{person}{Oscar Nierstrasz}
  {and} \bibinfo{person}{Michel Lemoine}} (Eds.), Vol.~\bibinfo{volume}{1687}.
  \bibinfo{publisher}{Springer}, \bibinfo{pages}{253--267}.
\newblock
\urldef\tempurl%
\url{https://doi.org/10.1007/3-540-48166-4\_16}
\showDOI{\tempurl}


\bibitem[\protect\citeauthoryear{Zeller and Hildebrandt}{Zeller and
  Hildebrandt}{2002}]%
        {DD}
\bibfield{author}{\bibinfo{person}{Andreas Zeller} {and} \bibinfo{person}{Ralf
  Hildebrandt}.} \bibinfo{year}{2002}\natexlab{}.
\newblock \showarticletitle{Simplifying and Isolating Failure-Inducing Input}.
\newblock \bibinfo{journal}{\emph{{IEEE} Trans. Software Eng.}}
  \bibinfo{volume}{28}, \bibinfo{number}{2} (\bibinfo{year}{2002}),
  \bibinfo{pages}{183--200}.
\newblock
\urldef\tempurl%
\url{https://doi.org/10.1109/32.988498}
\showDOI{\tempurl}


\bibitem[\protect\citeauthoryear{Zhang, Tian, Xu, Dong, Tan, and Sun}{Zhang
  et~al\mbox{.}}{2023}]%
        {DBLP:journals/corr/abs-2312-13064}
\bibfield{author}{\bibinfo{person}{Mengxiao Zhang}, \bibinfo{person}{Yongqiang
  Tian}, \bibinfo{person}{Zhenyang Xu}, \bibinfo{person}{Yiwen Dong},
  \bibinfo{person}{Shin~Hwei Tan}, {and} \bibinfo{person}{Chengnian Sun}.}
  \bibinfo{year}{2023}\natexlab{}.
\newblock \showarticletitle{Lampr: Boosting the Effectiveness of
  Language-Generic Program Reduction via Large Language Models}.
\newblock \bibinfo{journal}{\emph{CoRR}}  \bibinfo{volume}{abs/2312.13064}
  (\bibinfo{year}{2023}).
\newblock
\urldef\tempurl%
\url{https://doi.org/10.48550/ARXIV.2312.13064}
\showDOI{\tempurl}
\showeprint[arXiv]{2312.13064}


\bibitem[\protect\citeauthoryear{Zhang, Xu, Tian, Cheng, and Sun}{Zhang
  et~al\mbox{.}}{2024}]%
        {zhang2024deep}
\bibfield{author}{\bibinfo{person}{Mengxiao Zhang}, \bibinfo{person}{Zhenyang
  Xu}, \bibinfo{person}{Yongqiang Tian}, \bibinfo{person}{Xinru Cheng}, {and}
  \bibinfo{person}{Chengnian Sun}.} \bibinfo{year}{2024}\natexlab{}.
\newblock \showarticletitle{Deep Dive into Probabilistic Delta Debugging:
  Insights and Simplifications}.
\newblock \bibinfo{journal}{\emph{arXiv preprint arXiv:2408.04735}}
  (\bibinfo{year}{2024}).
\newblock


\bibitem[\protect\citeauthoryear{Zhou, Peng, Xie, Sun, Ji, Li, and Ding}{Zhou
  et~al\mbox{.}}{2022}]%
        {DBLP:journals/tsc/ZhouPXSJLD22}
\bibfield{author}{\bibinfo{person}{Xiang Zhou}, \bibinfo{person}{Xin Peng},
  \bibinfo{person}{Tao Xie}, \bibinfo{person}{Jun Sun}, \bibinfo{person}{Chao
  Ji}, \bibinfo{person}{Wenhai Li}, {and} \bibinfo{person}{Dan Ding}.}
  \bibinfo{year}{2022}\natexlab{}.
\newblock \showarticletitle{Delta Debugging Microservice Systems with Parallel
  Optimization}.
\newblock \bibinfo{journal}{\emph{{IEEE} Trans. Serv. Comput.}}
  \bibinfo{volume}{15}, \bibinfo{number}{1} (\bibinfo{year}{2022}),
  \bibinfo{pages}{16--29}.
\newblock
\urldef\tempurl%
\url{https://doi.org/10.1109/TSC.2019.2919823}
\showDOI{\tempurl}


\end{thebibliography}


\end{document}